\documentclass[useAMS,usenatbib,fleqn]{mnras}
\usepackage{amsmath}
\usepackage{amssymb}
\usepackage{aas_macros}
\usepackage{graphics,graphicx,hyperref}
\usepackage[usenames,dvipsnames]{xcolor}
\usepackage{fixmath}

\newcommand{\Mh}{{\hbox{$M_{\rm h}$}}}
\newcommand{\Mhi}{{\hbox{$M_{{\rm h},0}$}}}
\newcommand{\Eh}{{\hbox{$E_{\rm h}$}}}
\newcommand{\dEh}{{\hbox{$\dot E_{\rm h}$}}}
\newcommand{\dMh}{{\hbox{$\dot M_{\rm h}$}}}
\newcommand{\Rh}{{\hbox{$R_{\rm h}$}}}

\newcommand{\sfr}{{\hbox{$\Psi_\star$}}}

\newcommand{\G}{{\hbox{\rm G}}}
\newcommand{\vh}{{\hbox{$v_{\rm h}$}}}
\newcommand{\eagle}{{\sc eagle}}
\newcommand{\Hub}{\mathcal{H}}
\newcommand{\ikea}{$I\kappa\epsilon\alpha$}
\newcommand{\reference}{{\sc reference}}
\usepackage[utf8]{inputenc}
\title[Feedback-regulated galaxy formation]{The  $I\kappa\epsilon\alpha$ model of feedback-regulated galaxy formation}

\author[Sharma \& Theuns]
{Mahavir Sharma$^{\href{https://orcid.org/0000-0001-9927-5255}{\includegraphics[width=0.4cm]{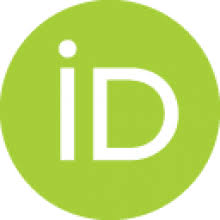}}1,2}$\thanks{mahavir.sharma@curtin.edu.au} and Tom Theuns$^{\href{https://orcid.org/0000-0002-3790-9520}{\includegraphics[width=0.4cm]{orcid.jpg}}3}$\\
$^{1}$International Centre for Radio Astronomy Research (ICRAR), Curtin University, Bentley, WA, 6102, Australia\\
$^{2}$ARC Centre of Excellence for All Sky Astrophysics in 3 Dimensions (ASTRO 3D), Australia\\
$^{3}$Institute for Computational Cosmology, Department of Physics, Durham University, Durham, DH1 3LE, UK
}
\begin{document}
\date{Submitted ---------- ; Accepted ----------; In original form ----------}
\maketitle
\begin{abstract}
We present the \ikea\ model of galaxy formation, in which a galaxy's star formation rate is set by the balance between energy injected by feedback from massive stars and energy lost by the deepening of the potential of its host dark matter halo due to cosmological accretion. Such a balance is secularly stable provided that the star formation rate increases with the pressure in the star forming gas. The \ikea\ model has four parameters that together control the feedback from star formation and the cosmological accretion rate onto a halo. \ikea\ reproduces accurately the star formation rate as a function of halo mass and redshift in the \eagle\ hydrodynamical simulation, even when all four parameters are held constant. It predicts the emergence of a star forming main sequence along which the specific star formation rate depends weakly on stellar mass with an amplitude that increases rapidly with redshift. We briefly discuss the emerging mass-metallicity relation, the evolution of the galaxy stellar mass function, and an extension of the model that includes feedback from active galactic nuclei (AGN). These self-regulation results are independent of the star formation law and the galaxy's gas content. Instead, star forming galaxies are shaped by the balance between stellar feedback and cosmological accretion, with accurately accounting for energy losses associated with feedback a crucial ingredient.
\end{abstract}
  \section{Introduction}
The cold dark matter cosmogony links the small fluctuations detected in the cosmic microwave background ({\sc cmb}) at redshift $z\sim 1000$ to  the observed large-scale clustering of galaxies at all observable redshifts. The fluctuations in the {\sc cmb} temperature correspond to density perturbations that grow in amplitude due to gravity, resulting in the formation of dark matter halos that host galaxies (see e.g. \citealt{Springel05a} and reference therein for more background). 

Whereas computer simulations can reliably predict virtually all properties of dark halos, the same can not be said for the properties of the galaxies that inhabit these halos. Even though our basic understanding of the underlying physics is probably correct  - galaxies form as gas accretes onto a halo, cools, becomes self-gravitating and forms stars \citep{White78, White91} - numerous uncertainties remain. What sets the star formation rate of a galaxy in a given halo at a given redshift? How does the energetic feedback from stars and accreting black holes regulate star formation? What is the role of galaxy interactions such as mergers? Are there any other crucial processes, for example feedback from cosmic rays or reionisation, and what is the role of magnetic fields?

Models that are designed to reproduce a mock universe that looks and evolves like the one we observe may not care about the details of the relevant physical processes. Examples include halo occupation distribution models ({\sc hod}, e.g. \citealt{Peacock00}) or subhalo abundance matching ({\sc sham}, e.g. \cite{Vale04}, see e.g. \cite{Wechsler18} for recent reviews).

Semi-analytical models recognise that the physics of galaxy formation is complex, and use parametrizations to model poorly understood physical processes. Cosmological hydrodynamical simulations try to capture some of these physical processes as accurately as possible (cosmological accretion and cooling of gas onto halos for example), but also rely on more parametrised descriptions of physical processes to capture physics below the resolution scale (see \citealt{Somerville15} and \citealt{Naab17} for recent reviews).

Several of the semi-analytical models and recent hydrodynamical simulations yield mock universe that look impressively similar to the one observed. Even though these models typically all include the same  ingredients, the details of how the processes are implemented may be quite different. It is therefore somewhat surprising that the resulting galaxy population is nevertheless very similar. At the very least this suggests some level of degeneracy in the modelling and that such calculations cannot be used to infer how the unresolved processes operate in detail. But it also suggests that many properties of galaxies do not actually depend on the details of many of these processes (see \citealt{Hopkins14} for a similar point of view).

Arguably one of the more striking features of the galaxy population as a whole is the emergence of a \lq star forming main sequence\rq\ (or \lq blue cloud\rq), \cite{Noeske07}, on which galaxies form stars at a specific rate, $\dot M_\star/M_\star$, that depends weakly on stellar mass ($M_\star$), but increases rapidly with redshift. The scatter around the mean trend is small, of order 0.3~dex (see \citealt{Schreiber15} and references therein for more recent observational analysis and discussion).

The appearance of such a main sequence suggests that the rate at which a galaxy forms stars in a halo of given mass, is somehow self-regulating. Several papers argued just that \citep[e.g.][]{Bouche10, Lilly13, Dekel14, Dave12, Dayal13}. The aim of these models is not to be able to predict the properties of galaxies in great detail, but rather understand the origin of self-regulation. The current paper follows this philosophy, adopting simplifications to more clearly expose the feedback loop that operates on the star forming sequence.

This paper is organised as follows: section~2 exposes the basic physics behind self-regulation in our model and tests the central assumptions by comparing to galaxies from the \eagle\ cosmological hydrodynamical simulation \citep{Schaye15}.  Section~3 explores consequences in terms of galaxy scaling relations (such as the galaxy stellar mass function and the mass-metallicity relation), compares these to simulations and data, and discusses successes and failures of the model.  Section~4 puts our results into context by comparing to previous work, and discusses what we think are its main limitations. Section~5 summarises our findings and is followed by an appendix that contains a short overview of the \eagle\ simulations, including a description of the reference model, \lq Ref-L100N1504\rq, in which the subgrid parameters are calibrated to reproduce redshift $z=0$ observations of the galaxy stellar mass function, the relation between galaxy size and mass, the relation between black hole mass and stellar mass, as described by \cite{Crain15}. The appendix also describes the \eagle\ model \lq FbConstNoAGN\rq, in which the feedback parameters are kept constant and which does not include feedback from AGN, as well as another \eagle\ variation,  \lq FbConst\rq,  in which the feedback parameters are kept constant and which does include AGN feedback.

\section{Self-regulation of star formation in galaxies}
The appearance of a star-forming sequence of galaxies is suggestive of the action of a feedback cycle. Such a feedback cycle is also important in understanding the main sequence of {\em stars} in a Hertzsprung-Russel diagram. Indeed: nuclear energy generation in main sequence stars is secularly stable - a prerequisite for their longevity. We begin this section by briefly describing the well-known reason behind this stability (see e.g. any text book on stellar structure, for example \citealt{Prialnik09}). We next investigate whether we can apply similar reasoning to star forming galaxies.


\subsection{The secular evolution of main sequence stars}
The total energy $E$ of a main sequence star of mass $M$ is the sum of its gravitational energy, $\Omega<0$, and its internal energy, $U=Mu$, where $u$ is its mean specific energy per unit mass. Stars are approximately in virial equilibrium, $E=\Omega/2=-Mu$, and as a consequence $dE/du<0$. Therefore, if a star loses energy for example through radiation so that $\dot E<0$, it will {\em heat up}, $\dot u>0$. The effective negative specific heat capacity of a star is a well-known but nevertheless intriguing feature of gravitationally bound systems, see e.g. \cite{Lynden-Bell77}, and is crucial for its longevity.

Indeed, consider a star losing energy through radiation (rate $L$), while gaining internal energy through nuclear fusion (rate $\dot E_{\rm nucl}$),
\begin{equation}
\dot E = \dot E_{\rm nucl} - L\,.
\label{eq:destar}
\end{equation}
In equilibrium, $\dot E=0$, however consider what happens for (small) deviations from equilibrium. Assuming $\dot E_{\rm nucl}<L$, say, $|E|$ increases since $E<0$, meaning $|u|$ increases and hence the temperature $T$ rises. The rate of energy generation through fusion is a rapidly increasing function of $T$, hence increasing $T$ increases $\dot E_{\rm nucl}$, so that $\dot E_{\rm nucl}<L$ results in an increase in $\dot E_{\rm nucl}$ towards equilibrium. Similarly, if $\dot E_{\rm nucl}>L$, the decrease in $T$ results in a decrease in the nuclear burning, until $\dot E_{\rm nucl}=L$.  Clearly, the negative specific heat capacity of a star is not just an amusing feature of self-gravitating systems, but is key in understanding stability on the main sequence. As the star's mean molecular weight changes due to fusion, $L$ and hence $\dot E_{\rm nucl}$ evolve secularly on a time scale which vastly exceeds $E/L$.

\subsection{The evolution of a galactic halo}
As a galactic halo\footnote{We will use the term \lq galactic halo\rq\ to refer to a central galaxy (as opposed to a satellite galaxy) with gas and stars, together with its host dark matter halo.} grows in mass due to cosmological accretion, its energy changes in time as well. At first sight there is little in common between the evolution of a galactic halo and that of a main sequence star. Indeed, the total energy of a star changes only secularly, $|\dot E|\ll L$, as self-regulation leads to a near balance between the energy generated by nuclear fusion and lost by radiation, but a galactic halo seems to have no equivalent channel for regulation. Does that mean that it is not secularly stable? The answer is partially yes: we show in the following that the dark matter halo is not secularly stable, in the sense that $\dot E_{\rm h} \neq 0$. However the same may not be true for the galaxy itself, because supernovae inject energy into the interstellar medium. Below we investigate whether that energy injection rate balances the loss of energy due to cosmological accretion, and if such a situation is a stable equilibrium - in analogy with the evolution of main sequence stars describe above. Before we do so we summarise some well known relations for the evolution of dark matter halos.

\begin{figure}
 \centering
 \includegraphics[width=\linewidth]{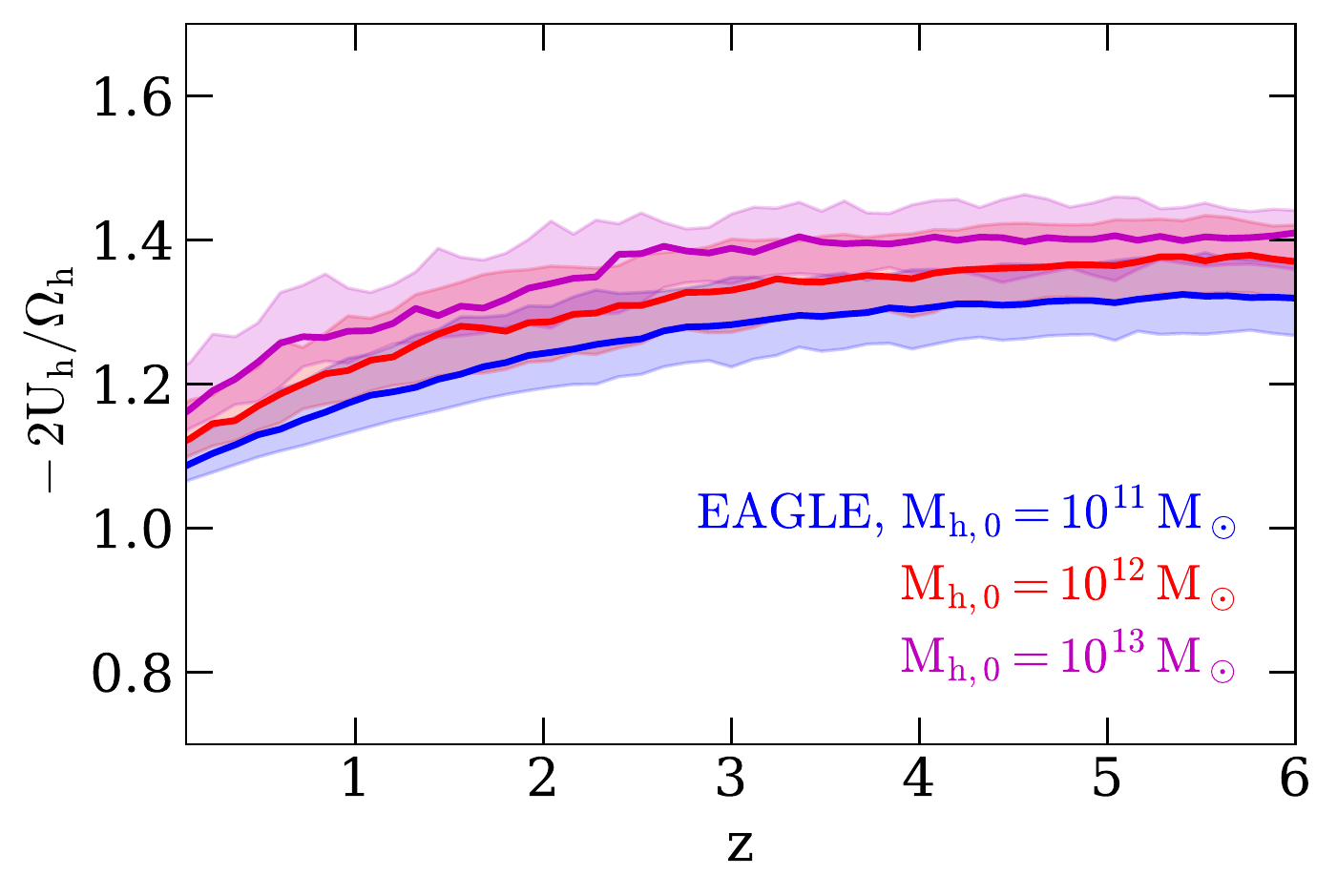}
 \caption{The redshift evolution of the virial ratio, $-2U_{\rm h}/\Omega_{\rm h}$, of dark matter halos from the \eagle\ L100N1504 dark matter only simulation tracked along their merger tree. Here, $U_{\rm h}$ is the sum of the kinetic energy of all particles in the centre of mass rest frame, and  $\Omega_{\rm h}$ is the gravitational energy.  Different colours refer to halos in narrow bins of their $z=0$ halo mass $M_{\rm h,0}$,  {\em blue}, {\em red} and {\em purple} correspond to $M_{\rm h, 0}=[0.98\hbox{--}1.02]\times10^{11}$~M$_\odot$, $[0.9\hbox{--}1.1]\times10^{13}$~M$_\odot$ and $[0.8\hbox{--}1.2]\times10^{13}$~M$_\odot$, respectively; solid curves are the median value of the virial ratio, the shaded region encompasses the 25th to 75th percentiles. Halos evolve approximately in virial equilibrium.}
\label{fig:VTz}
\end{figure}
\begin{figure}
 \centering
 \includegraphics[width=\linewidth]{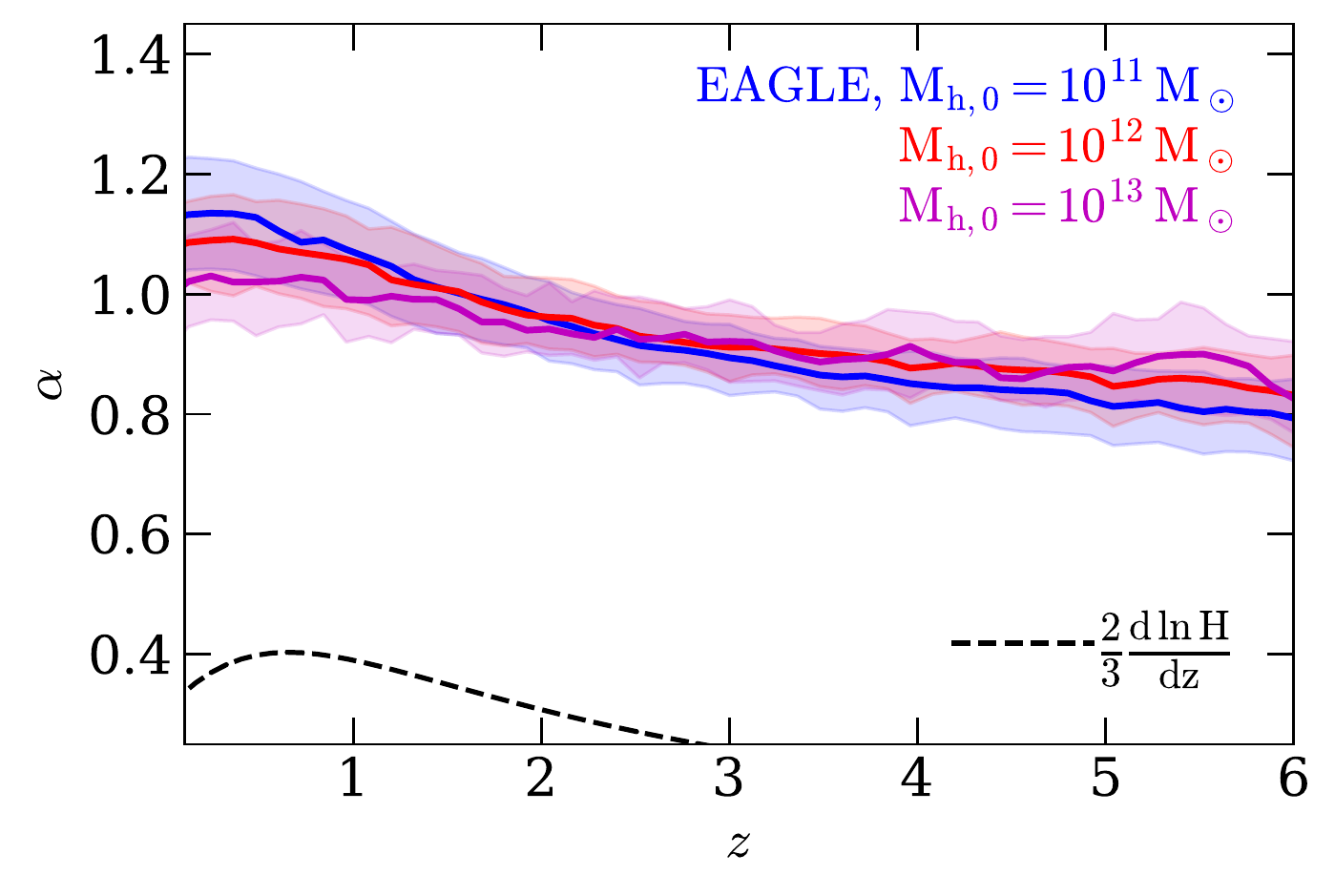}
 \caption{ As Fig.\ref{fig:VTz} but for the concentration parameter $\alpha$ from Eq.~(\ref{eq:halo}). As a halo grows, $\alpha$ remains approximately constant. The {\em dashed curve} quantifies the (negligible) effect of the last term on the right hand side in Eq.~(\ref{eq:de}).} 
 \label{fig:a1}
\end{figure}
\begin{figure}
 \centering
 \includegraphics[width=\linewidth]{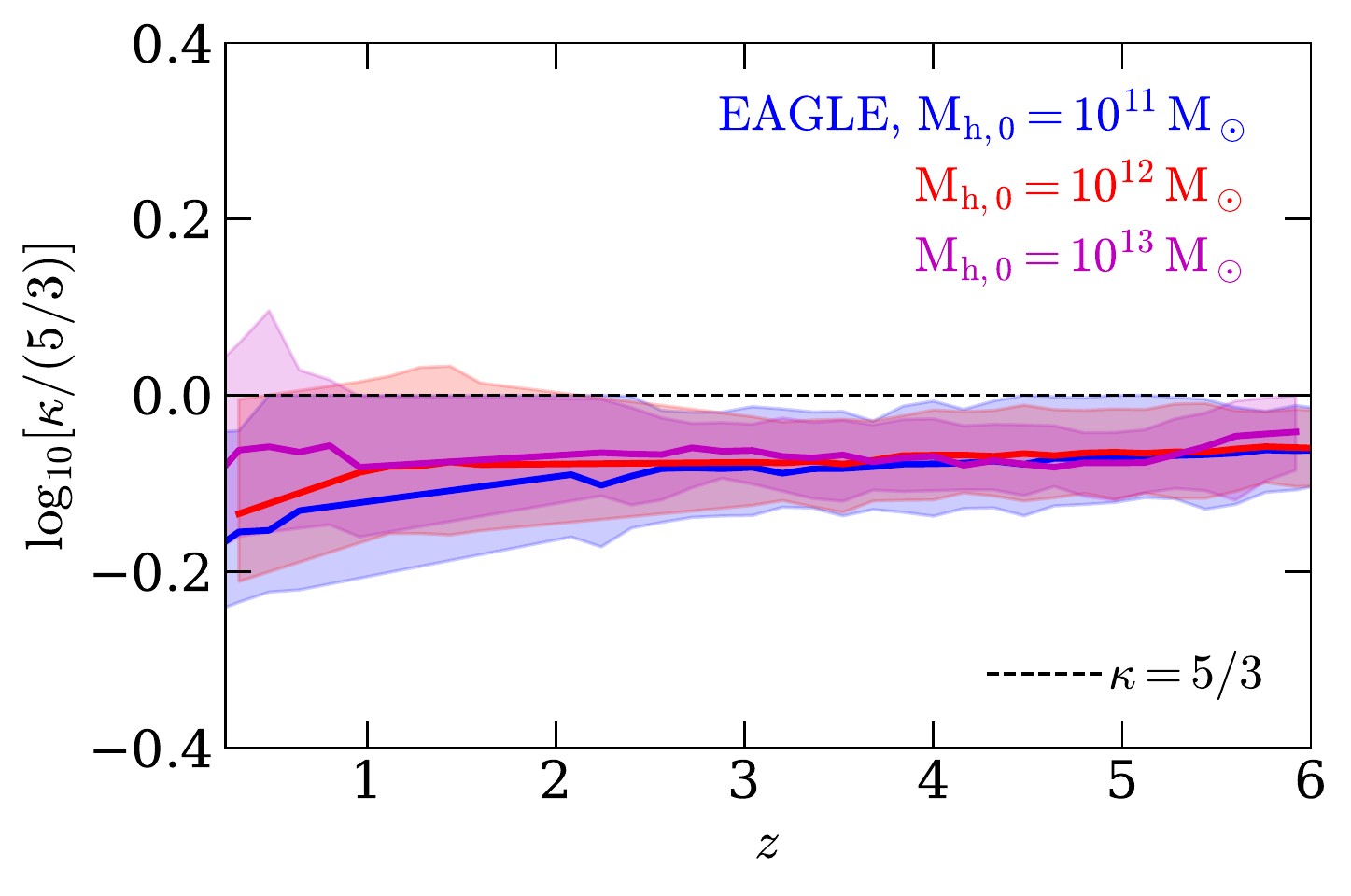}
 \caption{As Fig.\ref{fig:VTz} but for $\kappa={d(\ln|{E_{\rm h}}|)/ d(\ln{M_{\rm h}})}$, where $E_{\rm h}$ and $M_{\rm h}$ are the total energy and mass of a halo from Eq.~(\ref{eq:halo}).}
 \label{fig:k1}
\end{figure}

\subsection{The growth of a dark matter halo}
We begin by investigating the cosmological growth in mass and the associated change in energy of a dark matter halo. The concentration and mass of a dark matter halo may be affected by baryonic processes. Indeed, in the simulations presented by \citealt{Duffy10}, strong cooling and inefficient feedback increases the central dark matter density of galaxy and group halos significantly, whereas strong feedback, for example from an AGN, {\em decreases} that density. Baryonic mass loss, associated with strong feedback, may lead to a decrease in the rate at which a dark matter halo increases its mass. These effects are relatively modest at the scale of galaxies in the \eagle\ simulations, as shown by \citealt{Schaller15}, and we will neglect them in this paper.

The total energy, $E_{\rm h} <0$, of a dark matter halo with mass \Mh\ is the sum of its potential energy, $\Omega_{\rm h}<0$, and its internal energy, $U_{\rm h}$ (the total kinetic energy of all dark matter particles in the centre of mass rest frame, subscript $h$ for halo). Dark matter halos  satisfy the virial theorem approximately, $E_{\rm h}\approx {\Omega_{\rm h}/2} \approx -U_{\rm h}$ \citep[e.g.][]{Neto07}, as we show in Fig.~\ref{fig:VTz}. There is clearly some evolution of the ratio $U_{\rm h}/\Omega_{\rm h}$ as the halo grows, but we will neglect this in what follows. 

Assuming that the dark matter halo is in virial equilibrium, mass, radius and internal energy are related by,
\begin{eqnarray}
	\label{eq:halo}
	\Eh     &=& \Omega_{\rm h} + U_{\rm h} = {\Omega_{\rm h}\over 2} = -U_{\rm h}\nonumber\\
	\Omega_{\rm h} &=& -\alpha {\G\,M^{2}_{h}\over \Rh}\,\nonumber\\
    U_{\rm h}      &=& {1\over 2} \Mh\,\vh^2\,\nonumber\\
	\Rh    &=& \left({\G\Mh\over 100\,H^2}\right)^{1/3}\,.
\end{eqnarray}
We used the standard way of assigning a \lq radius\rq, \Rh, to a halo, by requiring that the mean density within \Rh\ is 200 times the critical density, $\rho_c=3H^2/(8\pi\G)$, where $H(z)$ is the Hubble constant at redshift $z$.  The value of the dimensionless parameter $\alpha$ depends on the halo's density profile: $\alpha=3/5$ for constant density, $\alpha = \Rh/(6a)$ for the spherical profile with scale radius $a$ described by \cite{Hernquist90}, and $\alpha$ is uniquely related  to the concentration parameter, $c$, of a halo with an NFW \citep{Navarro97} profile. Equations~(\ref{eq:halo}) also define a characteristic \lq virial velocity \rq\ of the halo, \vh, also given by
\begin{equation}
\vh^2=\alpha\,(\G\Mh)^{2/3}(10H)^{2/3}\,.
\label{eq:vh}
\end{equation}
If the accreting halo remains in virial equilibrium, then
\begin{equation}
{d\ln |\Eh|\over dz} = {5\over 3} {d\ln\Mh\over dz}+{d\ln\alpha\over dz}+{2\over 3}{d\ln H\over dz}\,.
\label{eq:de}
\end{equation}

We will show below that the first term on the right hand side, $|{5\over 3} {d\ln\Mh/ dz}|$ is of order unity. How about the other terms? We tracked the evolution of the parameter $\alpha$ of halos in the \eagle\ L0100N1504 dark matter only simulation along their merger tree. The result is plotted in Fig.~\ref{fig:a1}, where different colours refer to halos in bins of their redshift $z=0$ mass, $M_{\rm h,0}$. As was the case of the virial ratio, there is clearly some evolution in $\alpha$ as a halo grows, but that evolution is relatively weak and we will neglect it.
We also note that the term $(2/3)d\ln H/dz$ is always $<1/2$. Therefore the last two terms in Eq.~(\ref{eq:de}) are small compared to the first term on the right hand side, therefore ${d\ln |E_{\rm h}|/dz} \approx \kappa {d\ln\Mh/ dz}$ with $\kappa\approx 5/3$. To test this approximation in more detail, we once more track halos along their merger tree to compute $d\ln |\Eh|/d\ln\Mh$ directly, the result is plotted in Fig.~\ref{fig:k1}; different colours refer to halos in bins of $M_{\rm h,0}$. As $M_{\rm h}$ increases, $|E_{\rm h}|$ increases, with $d\ln |\Eh|/d\ln M_{\rm h}\approx 5/3$. Combining this approximation with  Eq.~(\ref{eq:vh}), motivates us to parametrize the rate of change of energy as a halo grows in mass by
\begin{equation}
\label{eq:dedt}
\dEh = -{\kappa\over 2}\,\dMh\vh^2\,.
\end{equation}
The variables $\alpha$ (Eq.~\ref{eq:vh}) and $\kappa$ are two of the four parameters of the \ikea\ model - and as we just showed, they are approximately independent of halo mass and redshift, and we will simply keep them constant at $\alpha=1$ and $\kappa=5/3$. We proceed by parametrizing the evolution of \Mh.

The increase with time of the halo mass in the extended  Press-Schecher (EPS) or  \lq excursion set\rq\ formalism \citep{Bond91, Lacey93} describes the growth measured in simulations very well. Here we will use the parametrisation described by \cite{Correa15a, Correa15b}, which we write in the form of the value of the halo mass at redshift $z=0$,
$\Mhi$, times a dimensionless function $m_{\rm h}(z)$,
\begin{eqnarray}
\Mh &=& \Mhi\, m_{\rm h}(z)\nonumber\\
m_{\rm h}(z) &\approx &(1+z)^{a}\exp(-bz)\,.
\label{eq:correaM}
\end{eqnarray}
The corresponding logarithmic growth rate  is
\begin{eqnarray}
{d\ln\Mh\over dz} &=& (a-b)\xi_{\rm h}(z)\nonumber\\
\xi_{\rm h}(z) &=& {1\over a-b}\left({a\over 1+z}-b\right)\,.\nonumber\\
\label{eq:correaMp}
\end{eqnarray}
The dimensionless functions $m_{\rm h}$ and $\xi_{\rm h}(z)$ are both unity at $z=0$. Since halos grow in mass, ${d\ln\Mh/dz}\le 0$, and in terms of the previous equation we have that the function $\xi_{\rm h}(z)>0$ but $a-b<0$. The parameters $a$ and $b$ depend on the mass of the halo at some reference redshift which we take to be $z=0$.  Averaging over halo masses, \cite{Correa15b} find
\begin{equation}
\bar a\approx 0.24\,,\quad \bar b\approx 0.75\,.
\label{eq:correaf}
\end{equation}
We will use $\bar a$ and $\bar b$ and denote them by $a$ and $b$ in our derivations, but in our figures we will use the more elaborate but more accurate version discussed by \cite{Correa15b} in which $a$ and $b$ are functions of $M_{\rm h,0}$ (except in Figure~\ref{fig:results} and \ref{fig:GSMF} in which we use the constant values from Eq.~\ref{eq:correaf}). Using $\bar a$ and $\bar b$, gives $|(5/3)d\ln\Mhi/dz|=0.85$ at $z=0$ and $1.25$ for $z\rightarrow\infty$, therefore $|(5/3)d\ln \Mh/dz|$ is of order unity, as we used before.

The virial velocity's evolution follows from Eq.~(\ref{eq:vh}),
\begin{equation}
\vh^2(z) = v_{\rm h,0}^2\,\left(m_{\rm h}(z)\Hub(z)\right)^{2/3}\,,
\label{eq:correavh}
\end{equation}
where the function $\Hub(z)$ is defined by
\begin{equation}
\Hub(z)\equiv {H(z)\over H_0}\,.
\label{eq:Ez}
\end{equation}

\subsection{The growth of a galaxy}
\subsubsection{Stability of feedback-regulated  galaxy formation}
\label{sect:selfreg}
\begin{figure}
 \centering
 \includegraphics[width=\linewidth]{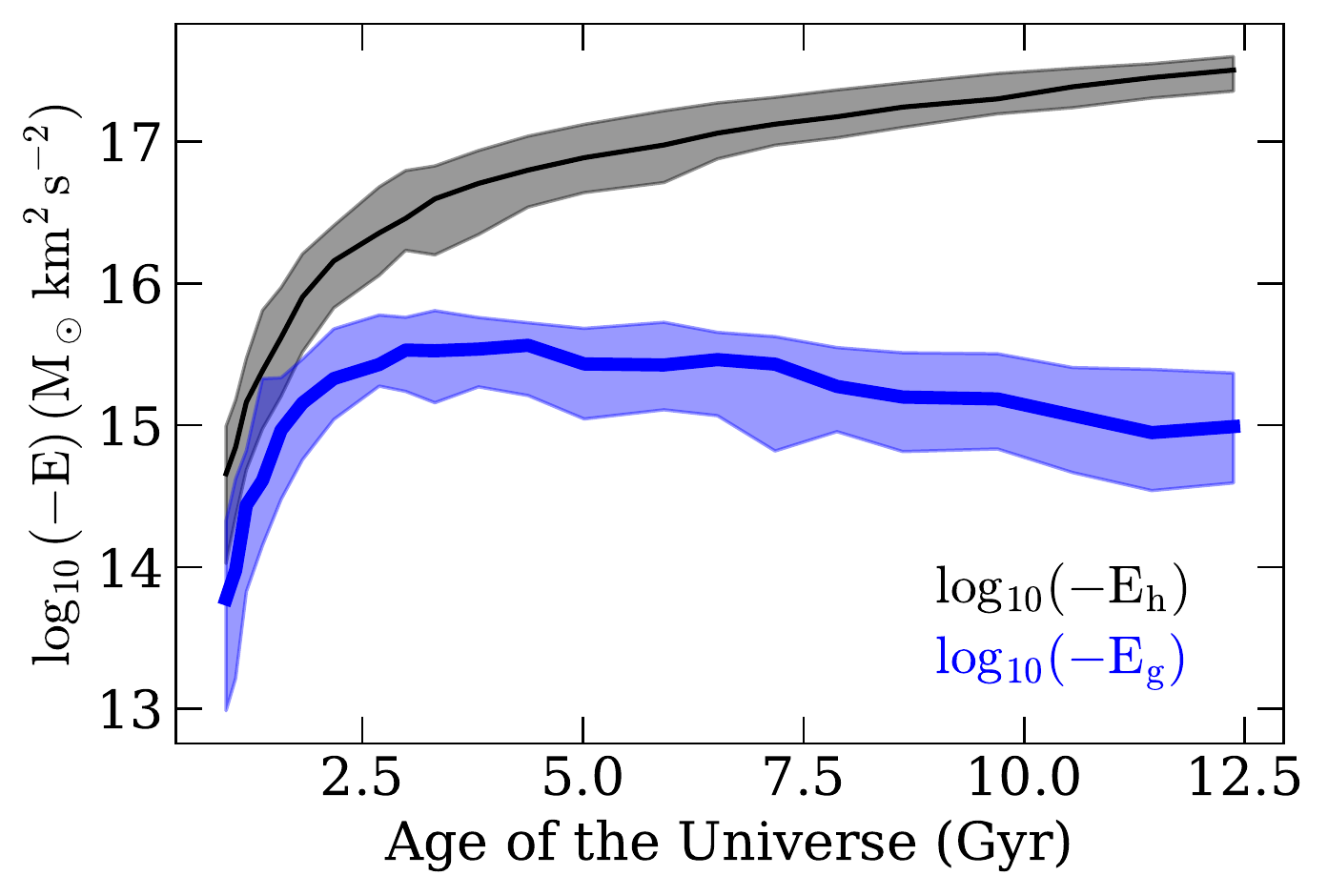}
 \caption{The evolution of the total energy of the dark matter halo, $E_{\rm h}$ ({\em black}) and the total energy of the star forming gas, $E_{\rm g}$ ({\em blue}) along the merger tree of a halo of $z=0$ mass, $M_{\rm h,0}=[0.8\hbox{--}1.2]\times 10^{13}~M_\odot$, selected from the \eagle\ simulation Ref-L100N1504. The solid curves show the median relation while the shaded area encompasses the 25$^{\rm th}$ to 75$^{\rm th}$ percentiles. While the total energy of the dark halo keeps on decreasing, the energy of the central galaxy decreases (secularly) at a slower rate as it is regulated by feedback from star formation.}
 \label{fig:Eg}
\end{figure}
A (central) galaxy too satisfies the equivalent of Eq.~(\ref{eq:dedt}). We neglect any pre-processing of the accreted matter, so that the ratio of gas mass that accretes onto the galaxy to total mass accreted onto the galactic halo, is simply the cosmological ratio $\omega_{\rm b}$ of the baryon to the total matter density, 
\begin{equation}
\omega_{\rm b}\equiv {\Omega_{\rm b}\over \Omega_{\rm dm}+\Omega_{\rm b}}={\Omega_{\rm b}\over\Omega_{\rm m}}\,.
\end{equation}
Once more neglecting the effect of the growing galaxy on the dark matter halo leads us to deduce that cosmological accretion decreases the energy of a galaxy at a rate $\dot E_{\rm g} =  -(\kappa/ 2)\omega_{\rm b}\dMh\vh^2$ (subscript \lq$g$\rq\ for galaxy).

However, unlike the case of the dark matter halo, the growing galaxy {\em generates} energy
through feedback from stars (and AGN, discussed later), therefore
\begin{equation}
\label{eq:edotG}
  \dot E_{\rm g} = \dot E_\star -{\kappa\over 2} \omega_{\rm b}\,\dMh\vh^2\,.
\end{equation}
In analogy with Eq.~(\ref{eq:destar}), we now speculate that $\dot E_\star\approx  {\kappa\over 2} \omega_{\rm b}\,\dMh\vh^2$: feedback from star formation compensates the energy loss associated with cosmological accretion so that the galaxy grows at nearly constant energy. Figure~\ref{fig:Eg} supports this {\em Ansatz}: it shows that, whereas the energy $E_{\rm h}$ of the dark matter halo (black curve) increases by almost 2 orders of magnitude from a look-back time of 10~Gyr to the present, the energy of the galaxy, $E_{\rm g}$, (blue curve) changes by less than $\sim 50$ per cent over the same time interval. 

Most of the energy injected into the galaxy's interstellar medium (ISM) is associated with star formation (i.e. supernovae and other processes associated with short-lived massive stars), therefore we write $\dot E_\star$ in terms of the star formation rate, $\dot M_\star$, and a characteristic velocity $v_\star$, 
\begin{equation}
  \dot E_\star = {1\over 2}\dot M_\star\,v_{\star}^2\,.
  \label{eq:estar}
\end{equation}
We can obtain an order of magnitude estimate for $v_\star$ by assuming that most of the injected energy is from core collapse supernovae (SNe), which inject $10^{51}$~erg of energy each and occur once per $100/\eta$ solar masses worth of stars formed\footnote{$\eta=1.74$ for a \citet{Chabrier03} stellar initial mass function that consists of stars in the mass range of $[0.1,100]{\rm M}_\odot$, of which those with mass $6-100~{\rm M}_\odot$ explode as a core collapse SN.}, hence
\begin{equation}
  \label{eq:sn}
  v_{\star} = \left(\epsilon\eta{2\times 10^{51}{\rm erg}\over 100~{\rm M}_\odot}\right)^{1/2} \approx 400 \left({\epsilon\over 0.091}\times {\eta\over 1.74}\right)^{1/2}~{\rm km~s}^{-1}\,.
\end{equation}
The  factor $\epsilon$ accounts for radiative loses, with $\epsilon=1$ corresponding to no radiative losses and $\epsilon\ll 1$ when such losses are
substantial.  Numerical simulations of SNe going off in a range of gas densities \citep[e.g.][and reference therein]{Thornton98}, and analytical models of the wind in M82 combined with simulations  \citep[e.g.][]{Strickland09}, suggest that a large fraction of the injected energy is radiated, $1-\epsilon\approx 90~\%$. The cooling rate of a radiating plasma also depends on its metallicity,  therefore $\epsilon$ is is unlikely to be constant in all galaxies and at all times. In this paper we use $\epsilon$ as a fitting parameter when comparing to the simulations; we used a reference value of $\epsilon=0.091$ in Eq.~(\ref{eq:sn}) which is consistent with the expected radiative losses being substantial {\em and} yields a round number for $v_\star$.

Equation~(\ref{eq:edotG}) that describes the rate of change of the energy of a galaxy is reminiscent of Eq.~(\ref{eq:destar}) that describes the rate of change of a main sequence star: whereas the star loses energy (becomes more bound) through radiative losses, the galaxy becomes more bound as the potential well of its host halo deepens due to cosmological accretion. While the star reacts by compensating the energy loss by nuclear fusion, the galaxy reacts by forming stars that inject energy in the galaxy's star forming gas. For stars, this results in 
$\dot E=\dot E_{\rm nucl}-L=0$, and we propose here that the same is true for a galaxy, $\dot E_{\rm g}=\dot E_\star-(\kappa/2)\omega_{\rm b}\dot M_{\rm h}\vh^2\approx 0$.

Why would the feedback from star formation be just so that $\dot E_{\rm g}\approx 0$? Is the equilibrium situation stable in the galaxy's case, just as it was for the star? To examine this question, suppose that $\dot E_{\rm g}<0$ - i.e. $|E_{\rm g}|$ is increasing because the galaxy is currently undergoing too little star formation given the current cosmological accretion rate. With gas in the galaxy getting compressed by the deepening potential well, the internal energy $U_{\rm g}$ of the galaxy will increase. How does that affect the star formation rate?

In the \eagle\ implementation of star formation, an increase in thermal energy per unit mass implies an increase in pressure, $P\propto u^4$ from Eq.~(\ref{eq:eos}), and hence an increase in star formation rate, $\dot\rho_\star\propto u^{4(n-1)/2}\approx u^{0.8}$ from Eq.~(\ref{eq:dotmstar1}) for $n=1.4$ from Eq.~(\ref{eq:KS}).  Therefore an increase in the accretion rate results in an increase in the star formation rate (and conversely, a decrease in the accretion rate results in a decrease in the star formation rate), so that the equilibrium\footnote{If the dynamical time-scales are very short, then self-regulation may fail to keep the galaxy in equilibrium. This may happen for example at high redshift,
\citep[e.g.][]{Duffy10}} situation, $\dot E_{\rm g}=0$, is secularly stable - just as in the case of nuclear fusion in a main sequence star, and for a similar reason\footnote{To take the analogy further, the galaxy in its galactic halo plays the role of the stellar core in the main sequence star.}. We note in particular that the increase in star formation rate due to increased accretion,  {\em neither} assumes {\em nor} requires that the gas mass - the fuel for star formation - increases. In our model, the gas reservoir is {\em not} regulating the star formation rate in a galaxy. We also note that stability requires that the star formation rate increases with the ISM's pressure, but without requiring any detailed form of the dependence of $\dot\rho_\star$ on $P$: {\em the details of exactly how star formation feedback operates are unimportant for the secular stability of the star formation rate in a star forming galaxy}. Another consequence is that the star formation rate in a cosmological galaxy depends very little, if at all, on the star formation law that relates star formation rate to the gas mass\footnote{We note this is not true in simulations of an {\em isolated} galaxy, for which the simulator specifies the gas fraction.}.

The star formation rate in our model of feedback-regulated galaxy formation, depends on the stellar {\bf  I}nitial mass function (through $\eta$ and the recycle fraction ${\cal R}$ discussed below) and the parameters $\mathbold{\kappa}$ (Eq.~\ref{eq:dedt}), $\mathbold{\epsilon}$ (Eq.~\ref{eq:sn}) and $\mathbold{\alpha}$ (Eq.~\ref{eq:halo}), which is why we called it \ikea. By computing the star formation rate and stellar mass as a function of halo mass, we next show that \ikea\ galaxies lie on a star-forming main sequence. 

\subsubsection{The main sequence of star forming galaxies}
\begin{figure}
	\centering
	\includegraphics[width=\linewidth]{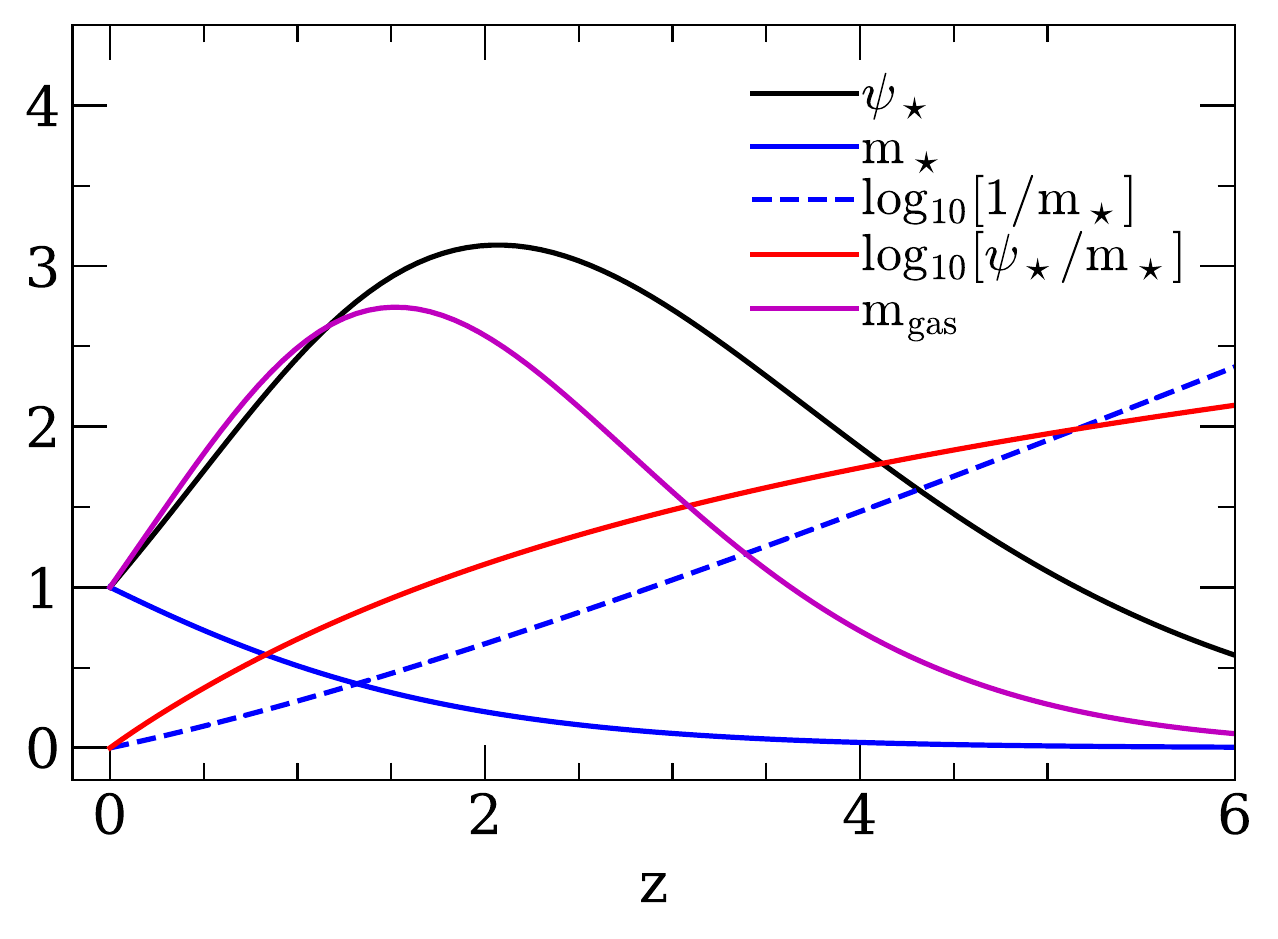}
	\caption{ The evolution of the dimensionless star formation rate $\Psi_\star(z)$ (black curve) from Eq.~(\ref{eq:dotmstar}), stellar mass $m_\star(z)$ (blue curve, we also plot $\log_{10}1/m_\star$ as a dashed blue line) from Eq.~(\ref{eq:mstar}), specific star formation rate $\sfr(z)/m_\star(z)$ (red curve) from Eq.~(\ref{eq:sSFR}), and the gas mass $m_{\rm gas}$ (magenta curve) from Eq.~(\ref{eq:Mgas}). We used $a=\bar a$ and $b=\bar b$ for the accretion history of halos, Eq.~(\ref{eq:correaMp}).}
\label{fig:results}
\end{figure}

Setting $\dot E_{\rm g}=0$ in Eq.~(\ref{eq:edotG}) for a self-regulating galaxy results in a relation between a galaxy's star formation rate and the cosmological accretion rate onto its host halo at a given redshift,
\begin{equation}
{1\over 2}\dot M_\star v_\star^2 = {\kappa\over 2}\omega_{\rm b}\dMh\,\vh^2\,,
\label{eq:SR}
\end{equation}
which is the main result of this paper. The right-hand side is the cosmological {\em energy} accretion rate onto a halo of given mass. The left-hand side sets the corresponding star formation rate in the galaxy, in terms of the effective energy injection rate per stellar mass formed. Substituting the expressions for the accretion rate $\dMh$ and virial velocity $\vh$ from Eqs.~(\ref{eq:correaM}) and (\ref{eq:correavh}) allows us to write the star formation rate as a product of its value at $z=0$, $\dot M_{\star, 0}$, times a dimensionless function, $\sfr(z)$, 
\begin{eqnarray}
\dot M_\star(z) &=& \kappa\omega_{\rm b}\dMh {\vh^2\over v_\star^2}  \equiv \dot M_{\star, 0}\sfr(z)\nonumber\\
\dot M_{\star, 0} &=& \kappa\omega_{\rm b}(b-a)H_0M_{\rm h,0}{v^2_{\rm h,0}\over v_\star^2}\nonumber\\
&=&1.2~{\rm M}_\odot{\rm yr}^{-1}{\kappa\over 5/3}{\alpha\over 1}\left[{h \over 0.677}\right]^{5/3}\left[{M_{\rm h,0}\over 10^{12}~{\rm M}_\odot}\right]^{5/3}\nonumber\\
&\times &\left[{v_\star\over 400~{\rm km~s}^{-1}}\right]^{-2}\nonumber\\
\sfr(z) &=& (1+z)\xi_{\rm h}(z)\left(m_{\rm h}(z)\,\Hub(z)\right)^{5/3}\,.
\label{eq:dotmstar}
\end{eqnarray}
The star formation rate scales $\propto M_{\rm h,0}^{5/3}\propto v_{\rm h,0}^5$; the function $\sfr(z=0)=1$.

Since stars lose mass during stellar evolution, the time integral of the star formation rate does not equal the total stellar mass at some later time. In the \lq instantaneous recycling approximation\rq, 
\begin{equation}
M_\star(t) = (1-{\cal R}) \int_0^t\dot M_\star(t')\,dt' \,,
\label{eq:Mstar}
\end{equation}
where ${\cal R}$ is the fraction of mass originally in stars that is returned back to star forming gas through stellar mass loss; the stellar population models used in \eagle\ have $1-{\cal R}\approx 0.55$ \citep{Wiersma09}. The stellar mass is in this approximation
\begin{eqnarray}
M_\star(z)  &=& (1-{\cal R}){\dot M_{\star,0}\over H_0}\,\int_z^\infty {\sfr(z') (1+z')^{-1}\mathcal{H}(z')}^{-1}\,dz'\nonumber\\
                 &\equiv & M_{\star,0}\,m_\star(z)\nonumber\\
M_{\star,0} &=& (1-{\cal R}){\dot M_{\star,0}\over H_0}\,m_{\star,0}\nonumber \\ 
                  &=&  1.7\times 10^{10}{\rm M}_\odot {1-{\cal R}\over 0.55}\left[{h\over 0.677}\right]^{2/3}
                            {\kappa\over 5/3}{\alpha\over 1}\nonumber\\
                            &\times & \left[{M_{\rm h,0}\over 10^{12}~{\rm M}_\odot}\right]^{5/3}
                  \left[{v_\star\over 400~{\rm km~s}^{-1}}\right]^{-2}
                   \nonumber\\
m_\star(z)  &=& {1\over m_{\star,0}}\int_z^\infty {\sfr(z') (1+z')^{-1}\mathcal{H}(z')}^{-1}\,dz'\nonumber\\
m_{\star,0} &=&  \int_0^\infty {\sfr(z') (1+z')^{-1}\mathcal{H}(z')}^{-1}\,dz'=1.78\,,
\label{eq:mstar}                            
\end{eqnarray}
with $m_\star(z=0)=1$.  To evaluate $M_{\star, 0}$ and $\dot M_{\star,0}$ we have used the cosmological parameters $\Omega_{\rm b}=0.0482519$, $\Omega_{\rm m}=0.307$, $\Omega_\Lambda=1-\Omega_{\rm m}$,  $\omega_{\rm b}=0.157$ and $h=0.677$ from \cite{Planck14}, and set $a=\bar a$ and $b=\bar b$ for the redshift dependence of the halo accretion rate from Eq.~(\ref{eq:correaf}); numerical values in our figures correspond to the more general accretion histories from \citet{Correa15b}, for which $a$ and $b$  depend on $M_{\rm h,0}$.

The specific star formation rate, ${\rm sSFR}$, follows from combining Eqs.~(\ref{eq:dotmstar}) and (\ref{eq:mstar}),
\begin{eqnarray}
{\rm sSFR}(z) &\equiv&  {\dot M_\star(z)\over M_\star(z)} ={H_0\over m_{\star,0}\,(1-{\cal R})}\,   {\sfr(z)\over m_\star(z)}\nonumber\\
 &=&{H_0\over m_{\star,0}(1-{\cal R})}\approx 0.07~{\rm Gyr}^{-1}~\hbox{\rm at $z=0$}\,.\nonumber\\
\label{eq:sSFR}
\end{eqnarray}
This ratio would depend on halo mass and hence also on $M_\star$ if ({\em i}) we had taken into account that the halo accretion rate depends on halo mass rather than using the average accretion rate from Eq.~(\ref{eq:correaf}) and ({\em ii}) if one or more of the \ikea\ parameters were to depend on halo mass.

The expression for the sSFR at $z=0$ from Eq.~(\ref{eq:sSFR}) looks suspiciously simple: what sets this numerical value? Tracing back the definitions of the dimensionless functions $\Hub(z)$ (Eq.~\ref{eq:Ez}), $\sfr(z)$ (Eq.~\ref{eq:dotmstar}) and $m_\star$ (Eq.~{\ref{eq:mstar}), we see that these {\em only} depend on cosmology and the growth rate of dark matter halos. Changing the growth rate will change the value of the numerical constant $m_{\star,0}$ in Eq.~(\ref{eq:mstar}). The only other \ikea\ parameter that sets the sSFR is ${\cal R}$, the recycled mass fraction, which depends on the IMF. Therefore the value of the sSFR at $z=0$ depends on cosmology (through the accretion history of halos), and on the fraction of mass returned to the ISM during stellar evolution, ${\cal R}$ - and nothing else. This is of course a consequence of assuming that none of the \ikea\ parameters evolve.

The dimensionless functions $\sfr(z)$, $m_\star(z)$, and $\sfr(z)/m_\star(z)$ provide the unique connection between the stellar properties of a galaxy and the properties of its host halo - they are plotted in Fig.~\ref{fig:results}. The star formation rate of an \ikea\ galaxy varies over a factor of $\sim 6$ between $z=0$ and $z=6$, peaking at $z\sim 2$, with half the stellar mass forming below $z\sim 1$. The sSFR increases rapidly with redshift, and is higher than its $z=0$ value by factors of 4.6, 13.7 and 30 at redshifts 1, 2 and 3, respectively.

To summarise: \ikea\ predicts a main sequence of star forming galaxies along which the specific star formation rate
does not depend on $M_\star$, provided the \ikea\ parameters themselves do not depend on halo mass. The value of this specific star formation rate increases rapidly with redshift.

\subsubsection{Comparison to \eagle}
\begin{figure}
	\centering
	\includegraphics[width=\linewidth]{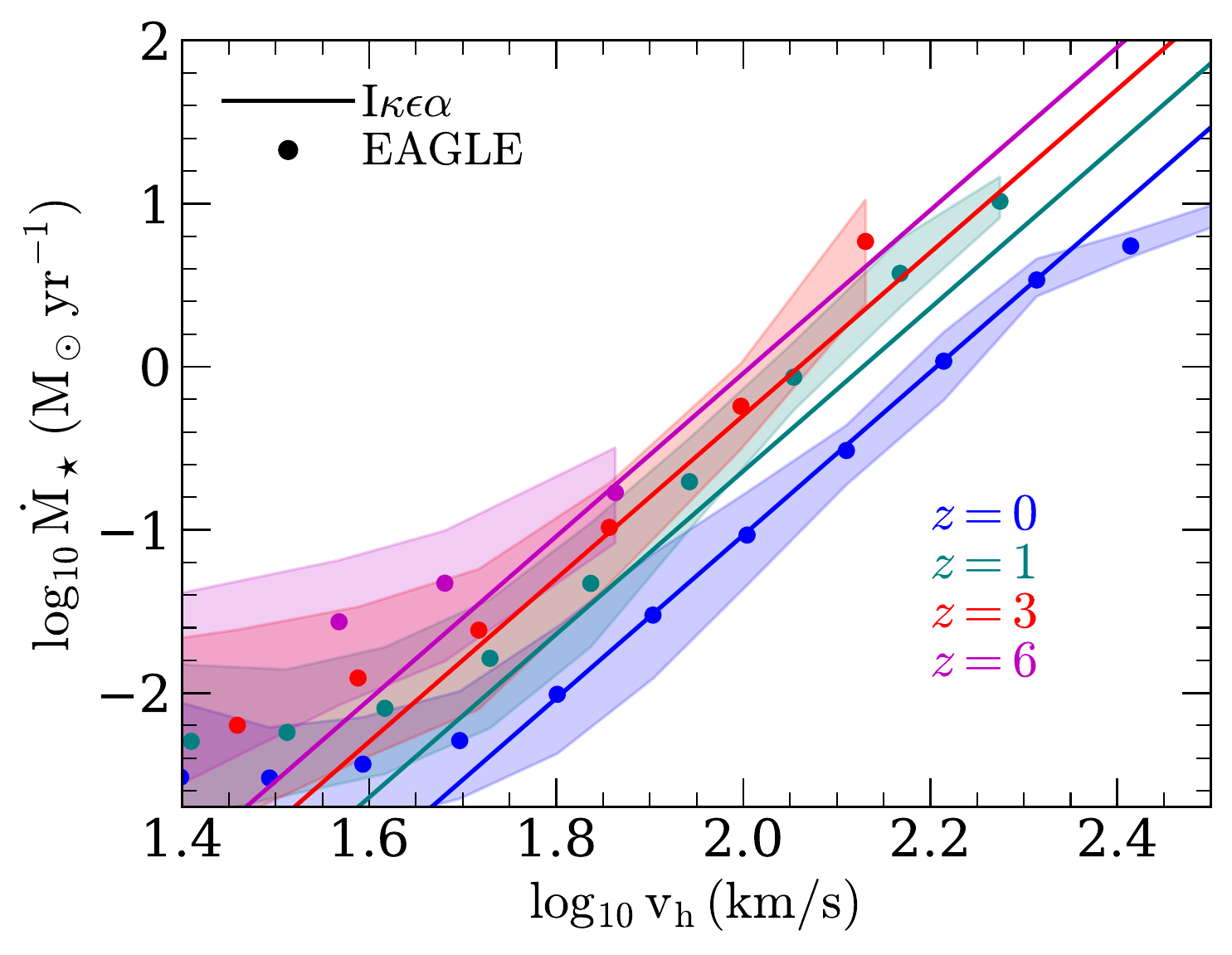}
	\caption{The dependence of the star formation rate, $\dot M_\star$, on the virial velocity \vh\ of a galaxy's host halo, at different redshifts.  The {\em coloured lines} are the predictions from the \ikea\  model (Eq.~\ref{eq:dotmstar}, with $\epsilon=0.2$) based on our self-regulation arguments;  {\em large dots} are the median star formation rate in \eagle\ galaxies (simulation FbconstnoAGN), with the {\em shaded area} encompassing the $25^{\rm th}-75^{\rm th}$ percentile range. Different colours correspond to different redshifts ({\em blue, green, red and purple} correspond to $z=0,1,3$ and 6, respectively). The \ikea\ model captures well the dependence of $\dot M_\star$ on \vh\ and $z$.}
	\label{fig:exp1}
\end{figure}
\begin{figure}
	\centering
	\includegraphics[width=\linewidth]{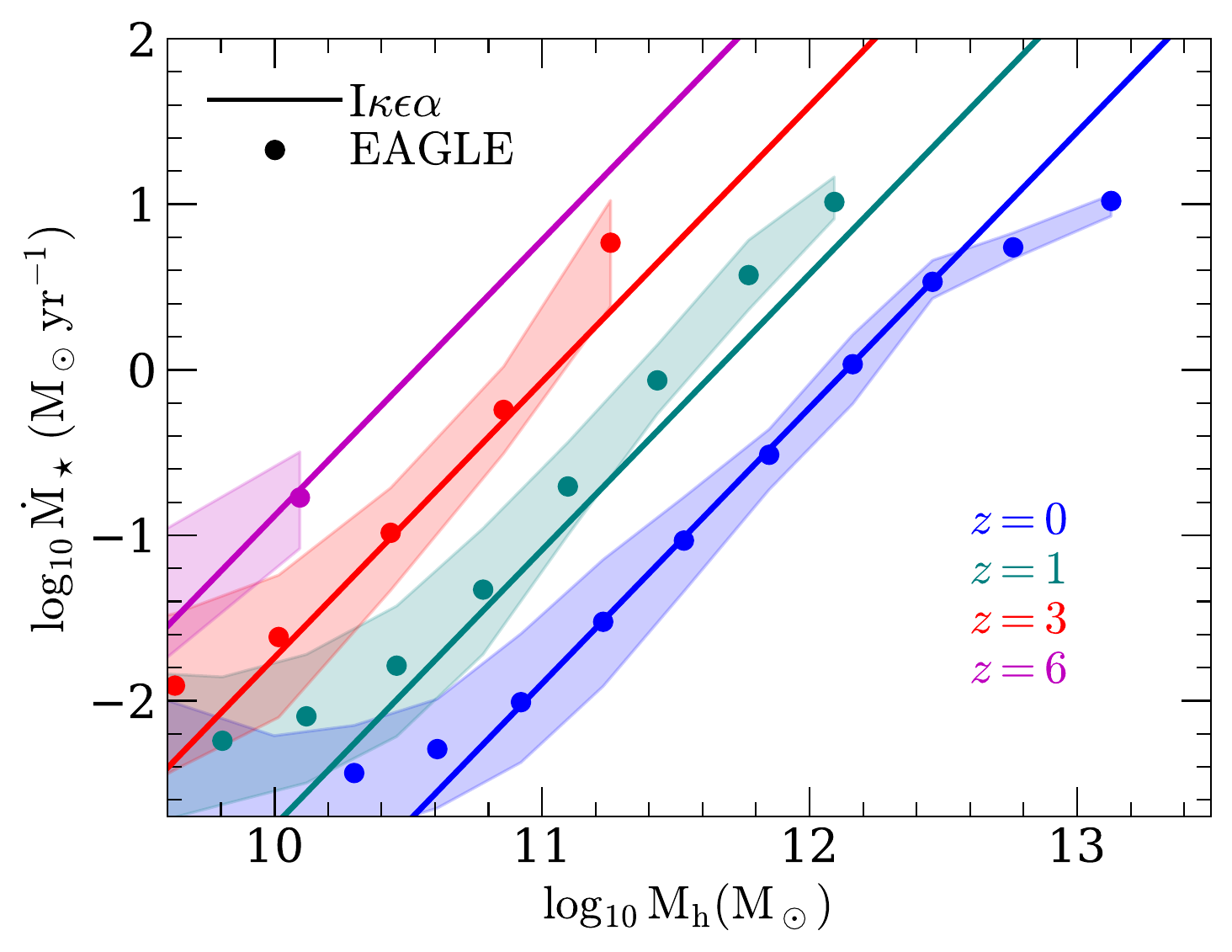}
	\caption{Same as Fig.~\ref{fig:exp1}, but for the dependence of $\dot M_\star$ on halo mass, \Mh. }
	\label{fig:exp2}
\end{figure}

\begin{figure}
	\centering
	\includegraphics[width=\linewidth]{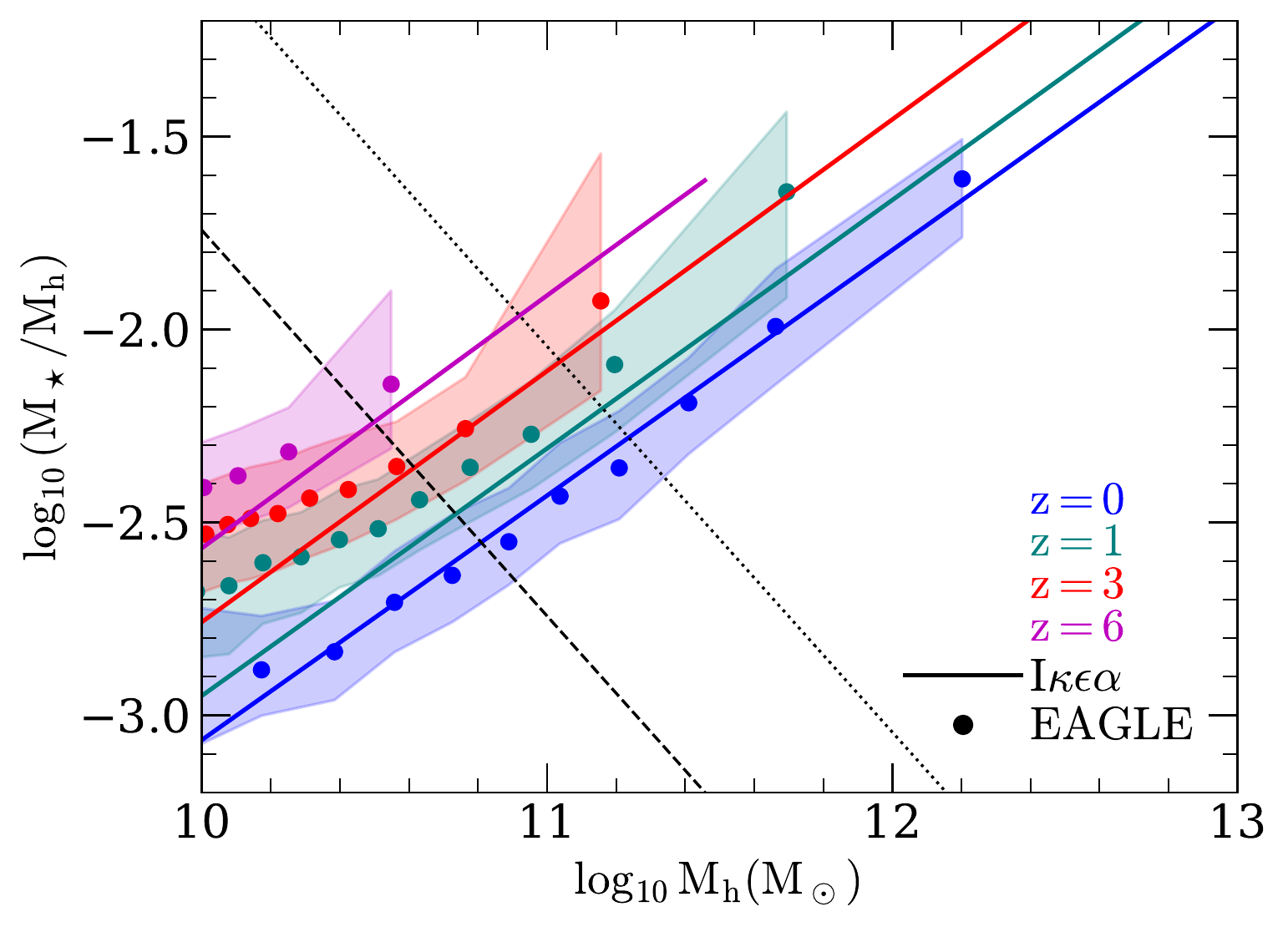}
	\caption{The stellar mass-halo mass ratio,  $M_\star/M_{\rm h}$, as a function of \Mh\ at different redshifts. The {\em coloured lines} are the predictions from the \ikea\ model ($M_\star$ from Eq.~\ref{eq:mstar}); {\em large dots} are the median relation in the \eagle\ galaxies (simulation FbconstnoAGN), with the {\em shaded area} encompassing the  $25^{\rm th}-75^{\rm th}$ percentile range. Different colours correspond to different redshifts ({\em blue, green, red and purple} correspond to $z=0,1,3$ and 6, respectively). The {\em black dashed} and {\em black dotted} lines correspond to \eagle\ galaxies with approximately 100 and 500 star particles, respectively. }
	\label{fig:SHMR}
\end{figure}

We test the ideas put forward in the previous section by comparing the star formation rate of galaxies as a function of halo properties and redshift to that of \eagle\ galaxies.  We emphasize that for a given assumed stellar IMF,  the $\epsilon$ parameter of the the \ikea\ model - a measure of the radiative loses in the ISM of the energy injected by SNe - is the central free parameter that sets the star formation rate in a cosmological halo. It does so by setting the characteristic velocity $v_\star$ through Eq.~(\ref{eq:sn}). The parameter $\epsilon$ likely depends on the properties of a galaxy's ISM - presumably $\epsilon$ would be smaller (greater cooling losses) when the ISM is denser and more metal rich. Rather than proposing a more detailed model for this, at this stage we simply keep $\epsilon$ constant. However, the \eagle\ reference simulation has a parameter $f_{\rm th}$ which explicitly changes the amount of energy injected into the ISM per solar mass of stars formed, depending on density and metallicity of the ISM (see Eq.~(7) in \citealt{Schaye15}).  Therefore, to keep the comparison between \ikea\ and \eagle\ fair,  we compare here to the \lq FBconstnoAGN\rq\ \eagle\ variation, in which $f_{\rm th}$ is kept constant (and which does not include AGN feedback either, see the Appendix for more details).  We re-iterate though, that keeping $f_{\rm th}$ constant is not quite equivalent to keeping $\epsilon$ constant, because the cooling losses in \eagle\ still depend on density and metallicity.

The star formation rate predicted by Eq.~(\ref{eq:dotmstar}) is compared to the \eagle\ FBconstnoAGN model in Fig.~\ref{fig:exp1}, taking  $\alpha=1$, $\kappa=5/3$, $\eta=1.7$ and $\epsilon=0.2$ (so that $v_\star$ is constant, see Eq.~\ref{eq:sn}); coloured lines are the \ikea\ prediction at different redshifts, large dots are the median relation for \eagle\ galaxies with the shaded region encompassing the $25^{\rm th}-75^{\rm th}$ percentile range. Even when keeping $v_\star$ constant, Eq.~(\ref{eq:dotmstar}) captures accurately the increase in $\dot M_\star$ with the halo's virial velocity \vh\ at fixed $z$, as well as the increase in $\dot M_\star$ with $z$ at fixed \vh. With only one \lq free\rq\ model parameter $\epsilon$ (which sets $v_\star$), we were astonished by the level of agreement between \ikea\ and \eagle.

In the case of Fig.~\ref{fig:exp1}, the increase in $\dot M_\star$ with $z$ at given \vh\ is due to the increase in the cosmological accretion rate onto a halo with given $\vh$ at given $z$, as is apparent from Eq.~(\ref{eq:SR}). However, plotting $\dot M_\star$ as a function of $\Mh$ (Fig.~\ref{fig:exp2}), we see that the redshift dependence is stronger due to the $\Hub(z)^{2/3}$ dependence of Eq.~(\ref{eq:dotmstar}). This is not surprising within the context of our self-regulation model:  the star formation rate depends on virial velocity rather than halo mass.

\subsection{The $M_\star-\Mh$ relation}
The stellar mass of a galaxy in Eq.~(\ref{eq:mstar}) is the product of a dimensional number that depends on the galaxy's halo mass at $z=0$, $M_{\rm h,0}$, times a dimensionless function $m_\star(z)$. This functional dependence allows us to answer the question of what is the $M_\star-\Mh$ relation in \ikea\ in  two different ways, ({\em i}) \lq What is the $M_\star-\Mh$ relation for a population of galaxies at a given redshift?\rq, and ({\em ii}) \lq How does the $M_\star-\Mh$ ratio of a halo evolve?\rq\ The answer to the first question follows from $M_\star\propto M_{\rm h,0}^{5/3}$ and $\Mh\propto M_{\rm h,0}$, therefore
\begin{equation}
{d\ln M_\star\over d\ln\Mh}|_{z={\rm const}} = {5\over 3}\,.
\end{equation}
The value of the exponent can be traced back to the $\dMh\,\vh^2\propto \Mh^{5/3}$ dependence of the star formation rate on halo mass, Eq.~(\ref{eq:dotmstar}). We compare the predicted relation to that measured in \eagle\ in Fig.~\ref{fig:SHMR}: coloured lines are the theoretical predictions at different redshifts, large dots are the median relation for \eagle\ galaxies with the shaded region encompassing the $25^{\rm th}-75^{\rm th}$ percentile range. Given that \ikea\ predicts the dependence of $\dot M_\star$ on \Mh\ as a function of $z$ in \eagle\ variation FBconstnoAGN well, it is not very surprising that it also reproduces the relation between $M_\star$ and \Mh.

Although \ikea\ galaxies lie along a line with $M_\star/\Mh\propto \Mh^{2/3}$, they do not evolve along this line. The $M_\star-\Mh$ ratio for a given halo evolves as
\begin{equation}
{d\ln M_\star\over d\ln\Mh}|_{M_{\rm h,0}={\rm const}}=-{m_{\rm h}(z)\over m_{\star,0}(a-b)m_\star(z)}(m_{\rm h}(z)\Hub(z))^{2/3}\,.
\end{equation}
This logarithmic slope is $\approx 1.1$ at $z=0$ and increases with $z$ to become nearly constant at a value of 1.4 for $z\ge 4$. If this slope were $5/3$, then (star forming) galaxies would evolve along the $z=0$ $M_\star-M_{\rm h}$ relation so that the stellar mass in a halo of a given mass would be independent of redshift. Because the slope is less than 5/3, the $M_\star/M_{\rm h}$ versus $M_{\rm h}$ relation evolves with redshift, in the sense that the stellar mass increases with redshift at a constant halo mass, however, that evolution is not very strong. This is the redshift evolution seen in Fig.~\ref{fig:SHMR}.

Summarising, we conclude that \ikea\ reproduces the relation between halo mass, star formation rate, and stellar mass measured in the FBconstnoAGN \eagle\ variation. The fact that \ikea\ reproduces the dependence of $\dot M_\star$ on \Mh\ is particularly encouraging, since it directly tests the very basis of the self-regulation argument of Eq.~(\ref{eq:SR}). Interestingly, the star formation rate predicted by Eq.~(\ref{eq:dotmstar}) does not depend {\em at all} on the galaxy's gas mass or indeed the assumed star formation law - as long as $\dot\rho_\star\propto u^\zeta$ for some sufficiently large and positive value of the exponent $\zeta$, so that the star formation rate {\em increases} with the pressure of the galaxy's interstellar medium. Instead the star formation rate depends on the cosmological accretion rate, and on $v_\star$ - that is, on the efficiency of stellar feedback.  We will return to this point in the discussion section. 

\subsection{The galaxy stellar mass function (GSMF)}
\label{sect:gsmf}
We compute the GSMF by combining the $M_\star-\Mh$ relation from \ikea\ with a model for the evolution of the halo mass function. The Press-Schechter (PS, \citealt{Press74}) approximation for the actual number density of halos per dex in halo mass \citep[e.g.][]{Reed07},  at $z=0$, is
\begin{equation}
{dn_{\rm h}\over d\log M_{\rm h,0}} = n_0\,({M_{\rm h,0}\over M_{\rm ps}})^{-\alpha_{\rm h}}\exp(-{M_{\rm h,0}\over M_{\rm ps}})\,,
\label{eq:hmf}
\end{equation}
where $n_0\approx 1\times 10^{-4}{\rm Mpc}^{-3}$ is a normalisation constant, $M_{\rm ps}\approx 2\times 10^{14}{\rm M}_\odot$ a characteristic mass above which the number density of halos falls exponentially, and the exponent $\alpha_{\rm h}\approx 0.9$. In the approximation that all halos grow at the same logarithmic rate, $\Mh(z)=M_{\rm h,0}\, m_{\rm h}(z)$, the halo mass function at redshift $z$ is 
\begin{equation}
{dn_{\rm h}\over d\log\Mh} = n_0\,({\Mh\over m_{\rm h}(z)\,M_{\rm ps}})^{-\alpha_{\rm h}}\exp(-{\Mh\over m_{\rm h}(z)M_{\rm ps}})\,,
\label{eq:PS}
\end{equation}
where $n_0$ and $M_{PS}$ are redshift independent, and $n_{\rm h}$ is now the co-moving number density of halos per dex in halo mass.  Provided the \ikea\ parameters are constants, $M_\star\propto \Mh^{5/3}$, and the co-moving number density of galaxies per dex in stellar mass becomes
\begin{eqnarray}
{dn_{\rm g}\over d\log M_\star}&=&{3n_0\over 5}
\left({M_\star\over m_\star(z)M_{\star, {\rm ps}}}\right)^{-\alpha_\star}\,\exp(-
{M_\star\over m_\star(z)M_{\star, {\rm ps}}})\nonumber\\
M_{\star, {\rm ps}} &=&M_{\star,0}\left({M_{\rm ps}\over 10^{12}{\rm M_\odot}}\right)^{5/3}\nonumber\\
&\approx&1.7\times 10^{10}\left({M_{\rm ps}\over 10^{12}{\rm M_\odot}}\right)^{5/3}{\rm M}_\odot\nonumber\\	
\alpha_\star &=& {3\over 5}\alpha_{\rm h}\approx 0.54\,.
\label{eq:gsmf}
\end{eqnarray}
In this approximation, the GSMF is just a scaled version of the halo mass function, with a power-law shape at low masses and an exponential cut-off at high masses. However it is well known that the \lq knee\rq\ in the galaxy stellar mass function - above which the exponential sets in - does not correspond to the knee in the halo mass function, but rather is a consequence of AGN feedback \citep{Bower06, Croton06}. We discuss how this can be incorporated in the model in \S \ref{sect:AGN} below.

It is interesting to note that we can make the same argument that lead to Eq.~(\ref{eq:gsmf}) to the star formation rate of a galaxy and compute the \lq galaxy star formation rate function\rq, GSRF, the number density of galaxies per dex of star formation rate.
Since $\dot M_\star \propto \Mh^{5/3}$, the GSRF has the same shape as the GSMF,
\begin{eqnarray}
{dn_{\rm g}\over d\log \dot M_\star}&=&{3n_0\over 5}
\left({\dot M_\star\over \Psi_\star(z)\dot M_{\star, {\rm ps}}}\right)^{-\alpha_\star}\,\exp(-
{\dot M_\star\over \Psi_\star(z)\dot M_{\star, {\rm ps}}})\nonumber\\
\dot M_{\star, {\rm ps}} &=&\dot M_{\star,0} \left({M_{\rm ps}\over 10^{12}{\rm M_\odot}}\right)^{5/3}\nonumber\\
&\approx&1.2\,\left({M_{\rm ps}\over 10^{12}{\rm M_\odot}}\right)^{5/3}{\rm M}_\odot~{\rm yr}^{-1}\,.
\label{eq:gsrf}
\end{eqnarray}
The constants $\dot M_{\star,0}(M_{\rm h,0}=10^{12}{\rm M}_\odot, z=0)$ and $M_{\star,0}(M_{\rm h,0}=10^{12}{\rm M}_\odot, z=0)$ are the $z=0$ star formation rate and stellar mass of a galaxy in a halo of mass $10^{12}{\rm M_\odot}$;
the numerical values for these are taken from Eqs.~(\ref{eq:dotmstar}) and (\ref{eq:mstar}), respectively.

At sufficiently low halo mass, these functions are power laws with slope $3\alpha_{\rm h}/5\approx 0.54$. The co-moving number density of galaxies with a given stellar mass increases monotonically with decreasing redshift $\propto m_\star(z)^{\alpha_\star}$. The corresponding evolution of the co-moving number density of galaxies with a given star formation rate is $\propto \Psi_\star(z)^{\alpha_\star}$: this function is not monotonic but peaks around $z\sim 2$. It falls to approximately $0.5$ and $0.54$ times its $z=2$ value at redshifts $z=5.5$ and $z=0$, respectively

The \ikea\ star formation in a halo with low $\vh$ is much less than the rate at which that halo accretes gas. Indeed, according to Eq.(\ref{eq:SR}), only a fraction $\kappa\vh^2/v_\star^2$ of the accreted gas goes into stars. What happens to the remaining gas? Also, self-regulation due to feedback from star formation must eventually fail for sufficiently high values of $\vh\approx v_\star/\kappa^{1/2}\approx 310~{\rm km}~{\rm s}^{-1}$, since then the star formation rate required to self-regulate would {\em exceed} the gas accretion rate. To investigate the consequence of these considerations  in more detail, we next examine the gas properties in \ikea.

\section{Galactic winds and the failure of self-regulating stellar feedback}

\subsection{Galaxy sizes and gas fractions}
\label{sect:outflows}
\begin{figure}
	\centering
	\includegraphics[width=\linewidth]{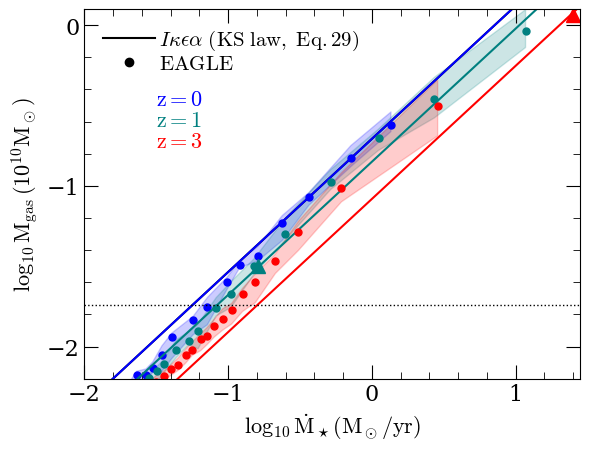}
	\caption{The mass of star forming gas in galaxies, $M_{\rm gas}$, versus the star formation rate, $\dot M_\star$, at different redshifts: $z=0$ (blue), $1$ (green) and $3$ (red). The {\em thick solid lines} show the scaling in \ikea\ obtained by using the Kennicut-Schmidt law (Eq.~\ref{eq:KSlaw}), with $n=1.4$ and assuming that the scale-length of the gas disk evolves as in Eq.~\ref{eq:rgas} \citep{Mo98}. {\em Large coloured dots} are the median relation in the \eagle\ galaxies (simulation FbconstnoAGN), with the {\em shaded regions} encompassing the  $25^{\rm th}-75^{\rm th}$ percentile range. {\em Dotted horizontal line} correspond to \eagle\ galaxies with 100 gas particles. The galaxies for which $R_{\rm gas}$ is less than the gravitational softening length in \eagle\ lie below 
		{\rm triangles} on each line, indicating that those galaxies are not well resolved in the \eagle\ simulation. In the \ikea\ model, the amount of gas in the interstellar medium is set by the star formation rate rather than the other way around.}
	\label{fig:Mgas}
\end{figure}
\begin{figure}
	\centering
	\includegraphics[width=\linewidth]{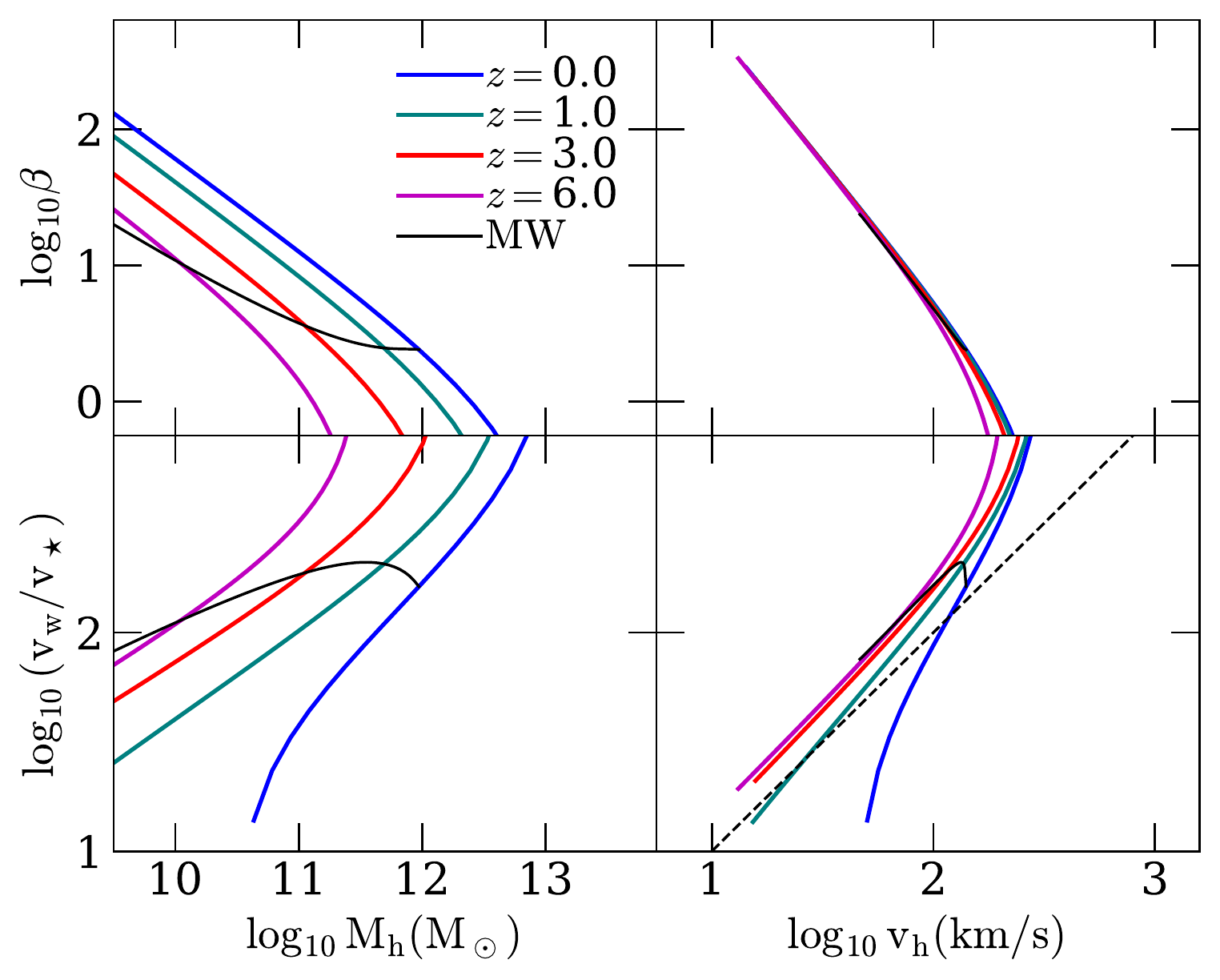}
	\caption{The dependence of the wind mass loading factor $\beta$ ({\em top panels}, from Eq.~\ref{eq:MB}) and the wind-speed, $v_{\rm w}$, at 5~times the wind launching radius ({\em bottom  panels}, from Eq.~\ref{eq:vw}) as a function of halo mass, \Mh ({\em left panels}), and halo virial velocity, \vh ({\em right panels}), from the \ikea\ model. The launching radius of the wind is taken to be equal to the gas scale radius, $R_{\rm gas}$ (Eq.\ref{eq:rgas}).  Coloured lines correspond to $\epsilon=0.1$, $\kappa=5/3$, $\alpha=1$, $1-{\cal R}=0.55$ in \ikea\ model, at different redshifts ({\em blue, green, red and purple} corresponding to $z=0$, 1, 3, and 6, respectively). The redshift dependence is stronger as a function of \Mh\ than as a function of \vh. At low values of $\vh\lessapprox 120$~km~s$^{-1}$, the wind speed tracks \vh, and the mass loading decreases from $\beta\sim 30$ at $\vh\sim 50$~km~s$^{-1}$ to $\beta\sim 1$ at $\vh\sim 100$~km~s$^{-1}$. The outflow begins to stall, $\beta\rightarrow 0$, for $\vh\rightarrow 180$~km~s$^{-1}$, at which point the wind speed becomes large, $\sim 10^3$~km~s$^{-1}$. The {\em black dashed line} in the bottom right panel is the one-to-one relation. The {\em thin black curve} labelled \lq MW\rq\ shows the evolution of $\beta$ and $v_{\rm w}$ for a Milky Way-like galaxy, $z=0$ halo mass of $M_{\rm h,0}=10^{12}{\rm M}_\odot$, as it grows in mass.}
	\label{fig:beta}
\end{figure}

\begin{figure}
	\centering
	\includegraphics[width=\linewidth]{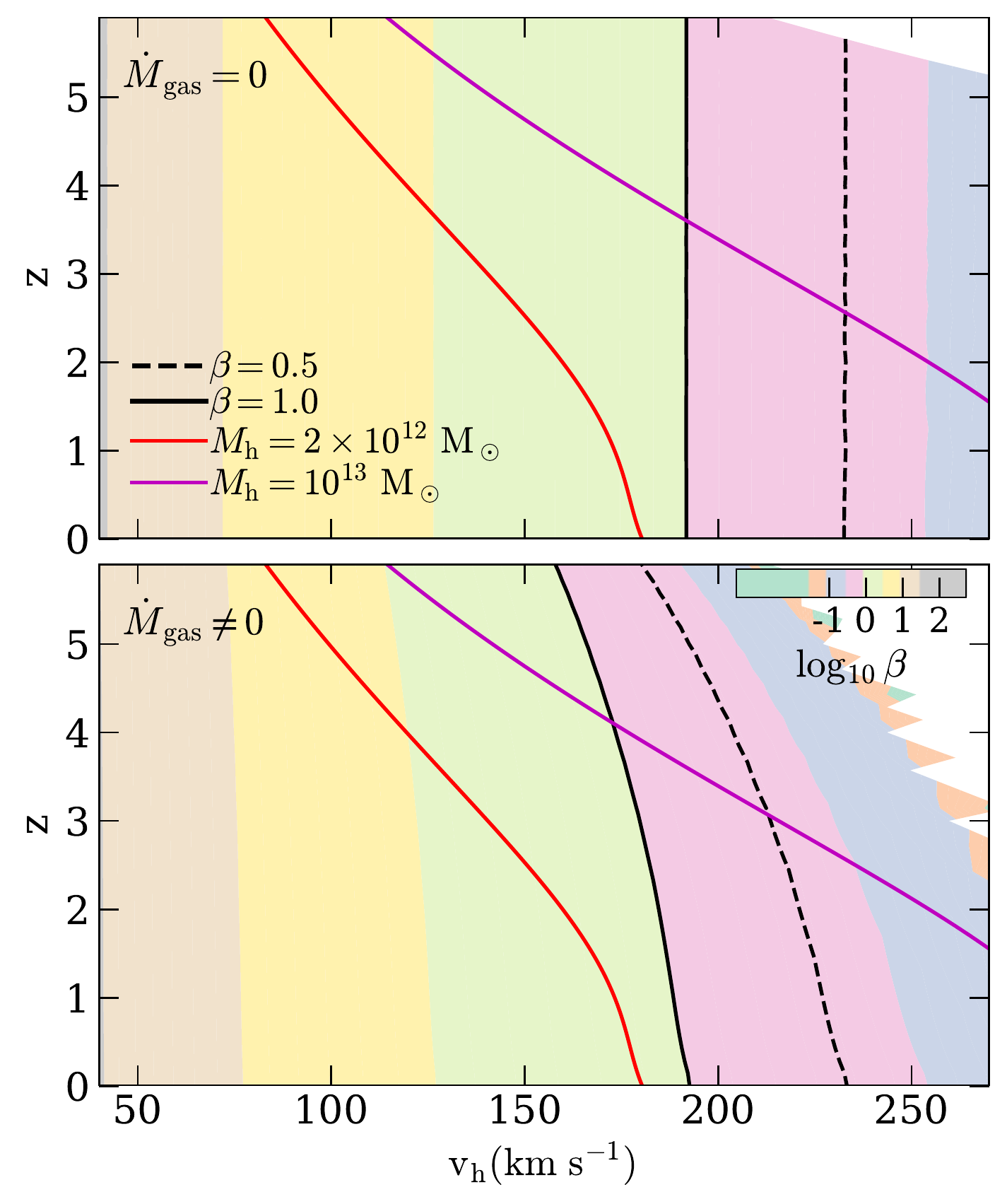}
	\caption{The mass loading factor $\beta$  from Eq.~(\ref{eq:MB2}) as a function of the halo virial velocity, \vh,  and redshift, $z$, shown as a color-map, for $\dot M_{\rm gas}=0$ ({\it top panel}) and for $\dot M_{\rm gas} \neq 0$ ({\it bottom panel}).  The dashed and solid black lines correspond to $\beta=0.5$ and $1$, respectively. }
	\label{fig:betamap}
\end{figure}

The star formation rate in the \ikea\ model does not depend on the gas mass. Instead, we {\em compute} the mass of star forming gas by assuming a star formation law. Taking the Kennicutt-Schmidt \citep{Kennicutt98} star formation law and assuming that star forming gas is in an exponential disk with scale-length $R_{\rm gas}$, the (total) star formation rate of a galaxy is related to its gas mass by
\begin{equation}
\dot M_\star = {2\pi A\,R_{\rm gas}^2\over n^2}\left({M_{\rm gas}/{\rm M}_\odot\over 2\pi (R_{\rm gas}/{\rm pc})^2}\right)^n\,,
\label{eq:KSlaw}
\end{equation}
where $A$ and $n$ are the parameters of the Kennicutt-Schmidt law (Eq.~\ref{eq:KS}).  We follow \cite{Mo98} 
(see also \citealt{Kravtsov13}) by assuming\footnote{\cite{Mo98} apply this reasoning to the stellar disk; \cite{Navarro17} show that $R_{\rm gas}$ scales better with the scale radius of the halo, but since we neglect variations in halo concentration by taking $\alpha=1$, these are equivalent.} that disk size, $R_{\rm gas}$,  is a constant fraction, $\lambda$, of the halo's virial radius. Using Eq.~(\ref{eq:halo}), this yields
\begin{eqnarray}
R_{\rm gas}&=&\lambda\,R_{\rm h}\, = R_{\rm gas, 0}\,r_{\rm gas}(z)\nonumber\\
R_{\rm gas, 0} &=& 2~{\rm kpc}\,{\lambda\over 0.01}\, \left({M_{\rm h,0}\over 10^{12}{\rm M}_\odot}\right)^{1/3}\nonumber\\
r_{\rm gas}(z) &=& {m_{\rm h}(z)^{1/3}\over \Hub(z)^{2/3}}\,,
	\label{eq:rgas}
\end{eqnarray}
where $\lambda=0.01$ yields a reasonable reference scale-length of $R_{\rm gas, 0}=2$~kpc for the galaxy inhabiting a $10^{12}{\rm M}_\odot$ halo at $z=0$. Using the $\dot M_\star-M_{\rm h,0}$ relation from Eq.~(\ref{eq:dotmstar}) and the $M_\star-M_{\rm h,0}$ relation from Eq.~(\ref{eq:mstar}) allows us to relate galaxy size to star formation rate and stellar mass, 
\begin{eqnarray}
{R_{\rm gas}(z)\over 2~{\rm kpc}} {0.01\over\lambda}&=&\left({\dot M_\star\over 1.2~{\rm M}_\odot~{\rm yr}^{-1}}\right)^{1/5}{r_{\rm gas}(z)\over \Psi_\star(z)^{1/5}}\nonumber\\
		&=&\left({M_\star\over 1.7\times 10^{10}~{\rm M}_\odot}\right)^{1/5}{r_{\rm gas}(z)\over m_\star(z)^{1/5}}\,.\nonumber\\
		\label{eq:mgms}
\end{eqnarray}
Sizes of a galaxies with a given $M_\star$ depend on redshift $\propto m_{\rm h}^{1/3}/(\Hub^{2/3}m_\star^{1/5})$. The ratio $m_{\rm h}^{1/3}/m_\star^{1/5}$ varies by less than a factor 0.75 below $z=6$, meaning that the size scales approximately as $1/\Hub(z)^{2/3}=1/(1+z)$ for $z\gg 1$, and slower than that at lower $z$. This agrees rather well with the observed scaling: \cite{Allen17} quotes a scaling $\propto (1+z)^{-0.97}$ for redshifts 5-7 and \cite{Vanderwel14} quotes a scaling $(1+z)^{-0.75}$
for redshifts 0-3. At a given value of $\dot M_\star$, $R_{\rm gas}\propto (1+z)^{-1.7}$ for $z\gg 1$, which is steeper than the $(1+z)^{-1.1}$ quoted by \cite{Shibuya15}. The weak dependence of size on mass, $R_{\rm gas}\propto M_\star^{1/5}$, is consistent 
the scaling $R_\star\propto M_\star^{0.22}$ for the stellar size-$M_\star$ relation found by \cite{Vanderwel14}.

Substituting Eq.~(\ref{eq:rgas}) into Eq.~(\ref{eq:KSlaw}) yields
\begin{eqnarray}
 M_{\rm gas} &=& M_{\rm gas,0}~m_{\rm gas}(z)\nonumber\\
 \left[{M_{\rm gas,0}\over 2.45\times 10^9~{\rm M}_\odot}\right]^n &=& {\kappa\over 5/3}{\alpha\over 1}\left[{h \over 0.677}\right]^{5/3}\,\left[{\lambda\over 0.01}\right]^{2n-2}\nonumber\\
&\times &\left[{v_\star\over 400~{\rm km~s}^{-1}}\right]^{-2}\left[{M_{\rm h,0}\over 10^{12}~{\rm M}_\odot}\right]^{1+2n/3}\nonumber\\
m_{\rm gas}(z) &=& [\sfr(z)]^{1/n}\,[r_{\rm gas}(z)]^{2-2/n}\,.
\label{eq:Mgas}
\end{eqnarray}
Using the $\dot M_\star-M_{\rm h,0}$ relation from Eq.~(\ref{eq:dotmstar}) allows us to compute the $M_{\rm gas}-\dot M_\star$ relation, and the result is compared to the \eagle\ simulation in Fig.~\ref{fig:Mgas}, where we used the values of $A$ and $n$ from \cite{Kennicutt98}. The \ikea\ prediction reproduces very well the slope of the relation and the normalisation at $z=0$. The simulated evolution is somewhat weaker than predicted. Although pleasing, the excellent agreement between the theoretical prediction and the simulation is not surprising: galaxies in \eagle\ follow the Kennicutt-Schmidt relation of Eq.~(\ref{eq:KS}), and that relation results in galaxies following Eq.~(\ref{eq:KSlaw}) at least approximately.

The evolution of the gas mass, as governed by the dimensionless function $m_{\rm gas}(z)$ from Eq.~(\ref{eq:Mgas}), is plotted in Fig.~\ref{fig:results}. The ratio of $m_{\rm gas}(z)$ over its value at $z=0$, is 0.5, 0.8, 1.2 and 1.4 at $z=4$, 3, 2 and 1, respectively, meaning that the gas mass of a forming galaxy changes by slightly more than a factor of 2 since $z=4$. Therefore assuming that galaxies form stars at nearly constant gas mass is a relatively good approximation below $z\sim 4$; it forms the basis of the equilibrium model of \cite{Dave12}, see also \cite{Bouche10,Krumholz12}.

\subsection{Galactic winds}
Galactic winds are a natural outcome of a model in which cosmological accretion sets the star formation rate but a star formation law sets the gas mass. Indeed, conservation of baryon mass requires that 
\begin{equation}
 \omega_{\rm b}\dMh = \dot M_{\rm gas} +(1-{\cal R})\dot M_\star +\dot M_{\rm w} \equiv \dot M_{\rm gas} +  (1+\beta-{\cal R})\,\dot M_\star\,,
\label{eq:MB}
\end{equation}
where $\dot M_{\rm w}$ is the rate at which the galaxy loses mass through a galactic wind, and the ratio $\beta\equiv \dot M_{\rm w}/\dot M_\star$ is usually called the \lq mass loading factor\rq\ of the wind. Solving for $\beta$ gives 
\begin{equation}
\beta =  {v_\star^2\over \kappa\vh^2} - (1-{\cal R}) - {\dot M_{\rm gas}\over \dot M_\star}\,,
\label{eq:MB2}
\end{equation}
where we used the main \ikea\ relation of Eq.~(\ref{eq:SR}) to relate $\dot M_\star$ and $\dot\Mh$. We can compute $\beta$ as a function of redshift and \vh\ or halo mass by integrating this equation using the relation between the gas mass and $\dot M_\star$ (Eq.~\ref{eq:KSlaw}), the result is shown in Figs.~\ref{fig:beta} and \ref{fig:betamap}. The velocity of the outflow can be estimated by assuming that the wind conserves energy once launched\footnote{This assumption may not be unreasonable because $\epsilon$ already accounts significant radiative losses before the wind is launched.},
\begin{eqnarray}
{1\over 2}\dot M_\star\,v_{\star}^2 = {1\over 2}\dot M_{\rm w}\,\left(1+{2\over {\cal M}^2\gamma(\gamma-1)}\right)v_{\rm w}^2 + {1\over 2}\dot M_{\rm w} v_\phi^2 \nonumber\\
{1\over 2}v_\phi^2 = \frac{c v_{\rm h}^2}{\ln(1+c) - {c \over 1+c}} \left[ {\ln(1+{c R_{\rm L} \over R_{\rm h}}) \over {c R_{\rm L}\over R_{\rm h}}} - {\ln(1+ {c R\over R_{\rm h}}) \over  {c R\over R_{\rm h}}} \right]\,. 
\label{eq:vphi} 
\end{eqnarray}
${\cal M}=v_{\rm w}/c_s$ is the wind's Mach number, $\gamma=5/3$ is the adiabatic index, $v_\phi^2/2$ is the change in potential of a Navarro-Frenk-White (NFW) halo \citep{Navarro97} between the launch cite, $R_L$, and the location $R$ where it is measured  \citep[e.g.][]{Lokas01}, $c$ is the halo's concentration parameter which depends on $M_{\rm h}$ and $z$ \citep[e.g.][]{Ludlow14} and we assume the launch radius, $R_{\rm L}=R_{\rm gas}$. $v_\phi\approx 0$ if the wind speed is measured very close to the launch site, and $v_\phi$ equals the escape speed from the halo if the wind speed is measured at infinity. This expression also neglects any ram-pressure the outflow may suffer. If the outflow is cold, ${\cal M}\rightarrow \infty$, and
\begin{equation}
\label{eq:vw}
v_{\rm w}^2 \approx \frac{v_{\star}^2}{\beta} - v_\phi^2\,.
\end{equation}
Clearly this treatment of the wind is quite approximate and in particular it is not obvious how one should compare our value of $v_{\rm w}$ to observations, in which the wind speed is often expressed in terms of the full width half maximum of an emission line. Fortunately, the behaviour of the mass loading $\beta$ is independent of these considerations, although here it is not so clear whether $\beta$ refers to gas leaving the galaxy or gas leaving the halo.

Given these limitations, we plot $\beta$,  and the wind speed, $v_{\rm w}$, at a distance of 5~times the gas scale radius, $R_{\rm gas}$, as a function of halo mass, virial velocity, and redshift in Fig.\ref{fig:beta}. The $\beta-\Mh$ relation evolves with redshift, as is clear from the left panels of the figure, basically because the relation between $\dot M_\star$ and \dMh\ depends on virial velocity according to Eq.~(\ref{eq:SR}). Most of that redshift dependence is removed if we plot $\beta$ as a function of \vh, as is seen from the right panels in the figure. As \vh\ increases, $\beta$ decreases and $v_{\rm w}$ increases. Also notice that as $\vh$ tends to a critical value of around $v_{\rm h,c}\approx 180$~km~s$^{-1}$, $\beta$ drops precipitously whereas the wind speed {\em increases} rapidly.

Winds in low \vh\ galaxies are slow and strongly mass-loaded, $\beta\gg 1$, as can be seen from Fig.~\ref{fig:beta}.  When $\beta\gg 1$ and making the further approximation that $|\dot M_{\rm gas}|\ll \dot M_\star$, Eqs~(\ref{eq:MB}) and (\ref{eq:vw}) combine to
\begin{equation}
v_{\rm w} = \left({\kappa (1+\beta-{\cal R})\over \beta}\right)^{1/2}\vh\approx \kappa^{1/2}\,\vh\,.
\end{equation}
Therefore, the wind speed tracks the halo's virial velocity (in low \vh\ galaxies at $z<4$), as is apparent from Fig.~\ref{fig:beta}. 

The relation between gas mass and star formation rate that results from the Kennicutt-Schmidt star formation law, Eq.~(\ref{eq:Mgas}), and the
equation for the mass-loading of winds, Eq.~(\ref{eq:MB2}), have interesting consequences, namely ({\em i}) the emergence of a mass-metallicity relation, and ({\em ii}) the existence of a characteristic value of \vh\ above which self-regulation due to feedback from stars fails. We investigate these next.

\subsection{The mass-metallicity relation}
\begin{figure}
	\centering
	\includegraphics[width=\linewidth]{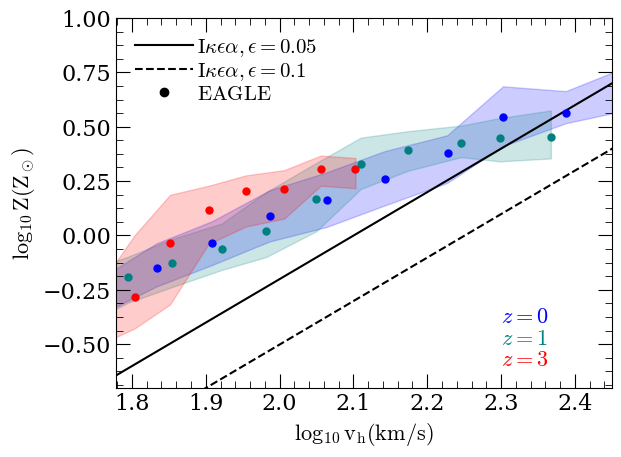}
	\caption{Metallicity, $Z$, of the star forming gas as function of the halo's circular velocity, \vh, at various redshift. Large solid dots
		are the median relation from \eagle\ (simulation FbconstnoAGN), with the shaded area encompassing the 25-75$^{\rm th}$ percentile for redshifts $z=0$ (blue), 1 (green) and 3 (red). Only halos with at least $10^3$ gas particles are shown. Black lines correspond to the \ikea\ model, from Eq.~(\ref{eq:ZMS}), with $\epsilon=0.05$ and 0.1 shown as a {\em solid} and {\em dashed line}, respectively; the redshift dependence of these lines is negligible, and the results depend very little on the assumed initial metallicity. The dependence of $Z$ on \vh\ is slightly shallower in \eagle\ compared to \ikea. The $Z-\vh$ relation is almost independent of redshift in both the model and \eagle.}
	\label{fig:Zvh}
\end{figure}
\begin{figure}
	\centering
	\includegraphics[width=\linewidth]{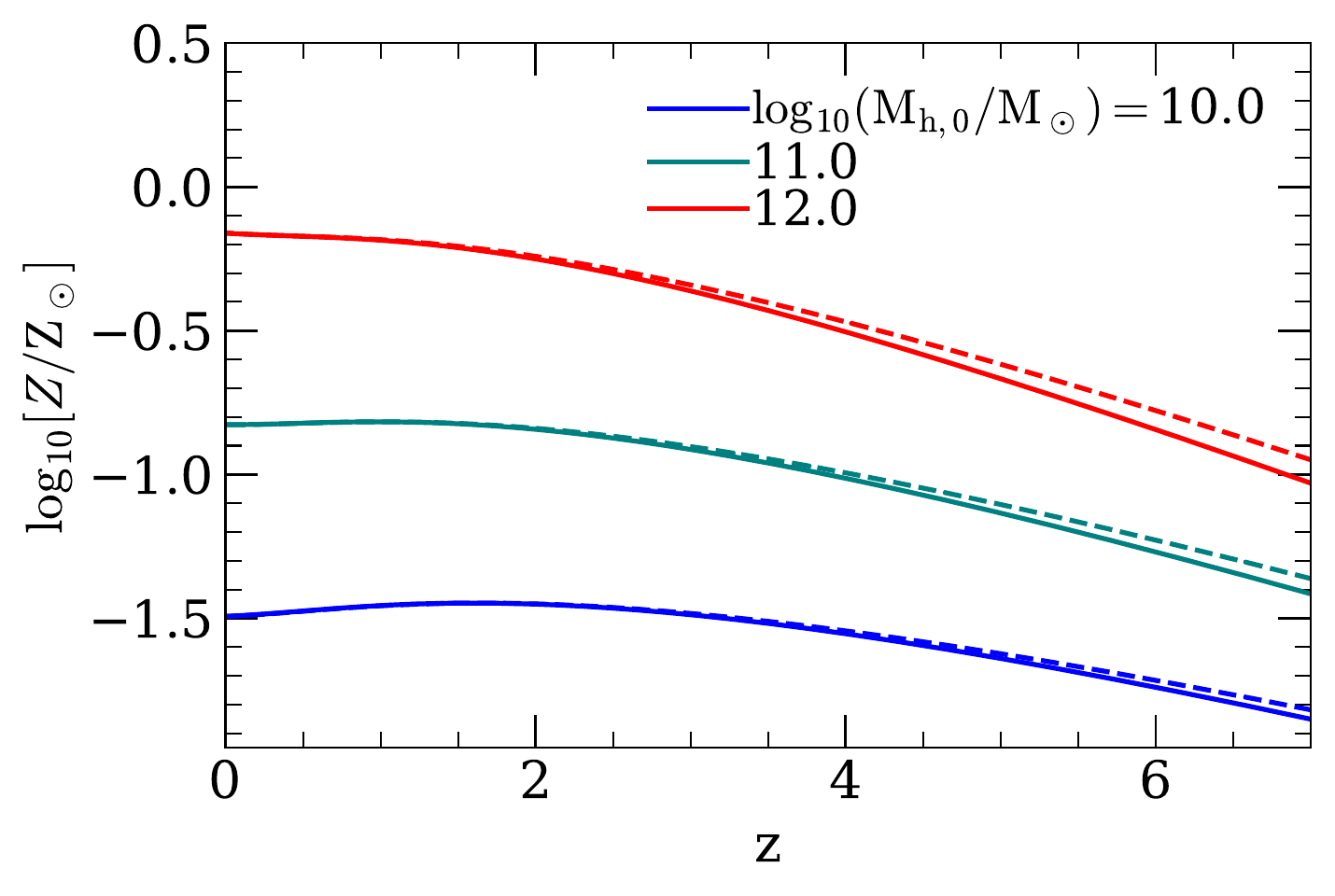}
	\caption{Metallicity, $Z$ in solar units, as a function of redshift, $z$, for halos with different mass; ({\em blue}, {\em green}, {\em red} correspond to $z=0$ halo masses of $\log_{10}M_{\rm h,0}/{\rm M}_\odot$=10, 11 and 12, respectively). The result from integrating Eq.~(\ref{eq:Zdot}) numerically is shown as {\em solid lines}, the approximation $Z=\kappa y\vh^2/v_\star^2$ is shown as {\em dashed lines}. Results are shown taking $v_\star=400$~km~s$^{-1}$, $y = 0.04$ and ${\rm Z_\odot}=0.0127$.}
	\label{fig:Zatt}
\end{figure}

The metal mass of the star forming gas, $M_Z\equiv Z\,M_{\rm gas}$, changes due to metals synthesised and released by stars, metals accreted, metals lost in a galactic wind, and metals locked-up in long-lived stars. 
Its rate of change is therefore
\begin{eqnarray}
  \dot M_Z &=& {d\over dt}(Z M_{\rm gas})\nonumber\\
  &=& y \dot M_\star + Z_0\omega_{\rm b}\dMh - Z_{\rm w}\dot M_{\rm w} - Z (1-{\cal R})
\dot M_\star\,,\nonumber\\
\label{eq:ZB}
\end{eqnarray}
where $y$ is the stellar yield, $Z_0$ is the metallicity of accreted gas and $Z_{\rm w}$ is the metallicity of
the wind which may differ from that of the gas, for example because enriched gas is more like to be ejected by feedback \citep[see e.g.][]{Creasey15}.
Combining this relation with Eq.~(\ref{eq:MB}), which expresses baryon mass conservation, and the main \ikea\ relation between $\dMh$ and $\dot M_\star$ from Eq.~(\ref{eq:SR}), we find that provided  $Z_{\rm w}=Z$ and $Z_0 = 0$, 

\begin{eqnarray}
\dot Z &=& y\frac{\dot M_\star}{M_{\rm gas}} - Z \frac{\omega_{\rm b} \dot M_{\rm h}}{M_{\rm gas}}\,, \nonumber\\
       &=& \frac{\dot M_\star}{M_{\rm gas}} \left( y - Z \frac{v_\star^2}{\kappa v_{\rm h}^2} \right)\,.
\label{eq:Zdot}
\end{eqnarray}
The recycled fraction ${\cal R}$ does not affect $\dot Z$ in the instantaneous recycling approximation, and the wind's mass loading $\beta$ does not affect $\dot Z$ provided $Z_{\rm w}=Z$. Integrating this equation in time,
we compare the relation between $Z$ and \vh\ as a function of redshift to the results from 
\eagle\ (simulation FbConstnoAGN) in Fig.~\ref{fig:Zvh}; the agreement is quite good, with \ikea\ showing a somewhat steeper dependence of $Z$ on \Mh\ and a lower normalisation at Milky Way-like values of $\vh\sim 140$~km~s$^{-1}$.
\begin{figure}
	\centering
	\includegraphics[width=\linewidth]{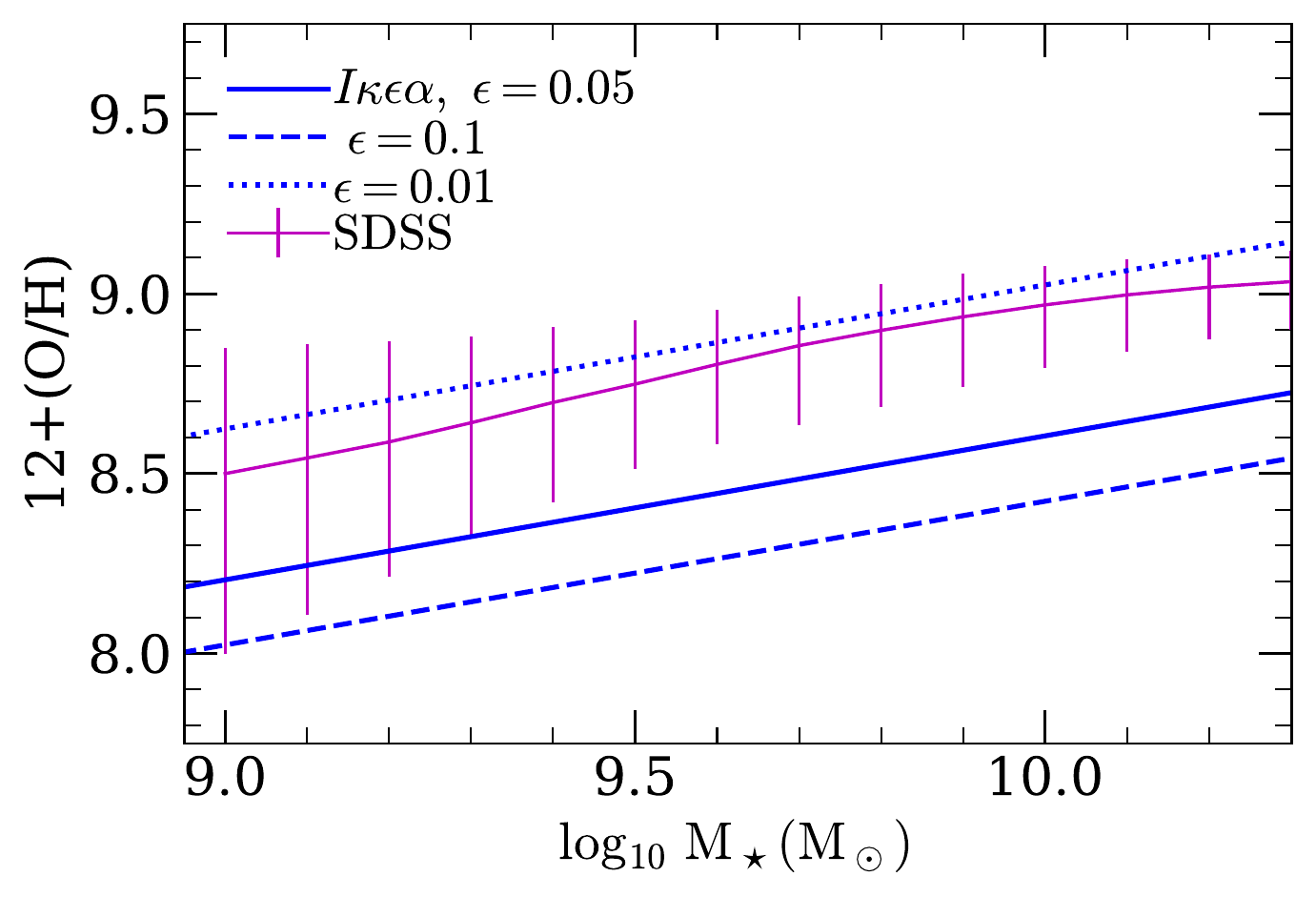}
	\caption{Metallicity 12+(O/H) as a function of the stellar mass, $M_\star$, from \ikea\ for feedback efficiency $\epsilon=0.01$ ({\em dotted blue}), $0.05$ ({\em solid blue}) and $0.1$ ({\em dashed blue}). The \ikea\ results (solid lines) have been obtained by integrating Eq.~(\ref{eq:Zdot}), for a yield, $y=0.04$ and then converted to the units of 12+(O/H) assuming that a metallicity of $Z=Z_\odot=0.0127$ corresponds to 12+(O/H)$=8.7$. For comparison the observed trend at $z=0$ for SDSS galaxies \citep{Mannucci10} is also shown as a thin magenta line with vertical bars for the scatter. \ikea\ reproduces the slope of the $Z-M_\star$ relation very well.}
	\label{fig:Zz}
\end{figure}

Interestingly though, both \ikea\ and \eagle\ show very little evolution of the $Z-\vh$ relation in Fig.~\ref{fig:Zvh}. Indeed, Eq.~(\ref{eq:Zdot}) shows that the metallicity of a galaxy tends to a value $Z\approx \kappa y\vh^2/v_\star^2$ that in fact only depends on a halo's virial velocity and not explicitly on redshift. In this approximation, the metallicity of a galaxy changes only secularly, tracking the evolution of $\vh^2$. Such a behaviour is an attractor of Eq.~(\ref{eq:Zdot}): given that $\dot M_\star/M_{\rm gas}>0$, $\dot Z$ is positive (negative) when $Z< y\kappa v_{\rm h}^2/v^2_\star$  
(when $Z> y\kappa v^2_{\rm h}/v^2_\star$). Therefore $Z$ approaches the secular value,
\begin{equation}
Z = \kappa y{v_{\rm h}^2\over v_\star^2}\,,
\end{equation}
on the gas consumption time-scale, $M_{\rm gas}/\dot M_\star$. This secular value reproduces the evolution from Eq.~(\ref{eq:Zdot}) very well, as shown by the dashed lines in Fig.~\ref{fig:Zatt}. 

Using this secular expression for $Z(\vh)$, taking $v_\star=400$~km~s$^{-1}$,  and $y=0.04$, $Z_\odot=0.0127$ for the total metal yield and solar metallicity as done in the 
the stellar evolution models collected from the literature by \citet{Wiersma09}, we obtain
\begin{eqnarray}
{Z\over 0.68\times Z_\odot} &=& \left(\frac{M_{\rm h,0}}{10^{12}~{\rm M_\odot}}\right)^{2/3}   (m_{\rm h}\Hub)^{2/3} \nonumber\\
&=& 
\left({\dot M_{\star}\over 1.2~{\rm M}_\odot{\rm yr}^{-1}}\right)^{2/5}
{(m_{\rm h}\Hub)^{2/3}\over \Psi_\star^{2/5}}\nonumber\\		
&=& 
\left({M_{\star}\over 1.7\times 10^{10}~{\rm M}_\odot}\right)^{2/5}
{(m_{\rm h}\Hub)^{2/3}\over m_\star^{2/5}}\,.
\label{eq:ZMS}
\end{eqnarray}
The reference values of $\dot M_\star$ and $M_\star$ for the star formation rate and stellar mass, are taken from Eqs.~(\ref{eq:dotmstar}) and (\ref{eq:mstar}). The normalisation of this relation, $0.68Z_\odot$ for $M_{\rm h,0}=10^{12}{\rm M}_\odot$, depends on \ikea\ parameters $\propto \left({\kappa\alpha/(\epsilon\eta)}\right)^{3/5}$. The observed normalisation is uncertain but at face value higher than what we find by a factor of two \cite[e.g.][]{Tremonti04}. 

The dependence of $Z$ on $\vh$ implies that \ikea\ galaxies fall on a mass-metallicity relation, as well as on a star-formation rate-metallicity relation. Similarly to the stellar mass-halo mass relation, we can compute how $Z$ depends on $M_\star$ at a given redshift,
\begin{equation}
{d\ln Z\over d\ln M_\star}|_{z={\rm const}}={d\ln Z\over d\ln \dot M_\star}|_{z={\rm const}}={2\over 5}\,,
\end{equation}
independent of redshift, with the value of the exponent resulting from the $\vh^2\propto M_\star^{2/5}$ dependence of Eq.~(\ref{eq:mstar}). As a galaxy grows in mass, its metallicity increases as
\begin{equation}
{d\ln Z\over d\ln M_\star}|_{M_{\rm h,0}={\rm const}}={2\over 3}{d\ln (m_{\rm h}\Hub)/dz\over d\ln m_\star/dz}\,.
\end{equation}

The evolution of $Z$ at a {\em given} stellar mass or star formation rate, is $\propto (m_{\rm h}\Hub)^{2/3}$ according to Eq.~(\ref{eq:ZMS}).
With increasing $z$, $m_{\rm h}(z)$ decreases whereas $\Hub(z)$ increases, resulting in little evolution in the $Z-M_\star$ relation. At a given value of $M_\star$, $Z$ decreases with increasing $z$ by factors 0.9 and 0.76 compared to its $z=0$ value for $z=2$ and $3$, respectively. The observed evolution is somewhat stronger and better reproduced by the \eagle\ {\sc reference} model in which $\epsilon$ varies with the local gas properties \citep{DeRossi17}.

Why does $Z$ depend on $M_\star$ in \ikea? The \ikea\ metallicity of a galaxy is $Z\approx y\dot M_\star/(\omega_{\rm b} \dot M_{\rm h})$, the ratio
of the rate at which stars metal enrich the ISM over  the rate at which these metals are being diluted by accreting (primordial) gas. The reason this ratio depends on $M_\star$ is that the star formation efficiency
depends on $v_{\rm h}$: $\dot M_\star/ (\omega_{\rm b}\dot M_{\rm h})\propto v_{\rm h}^{2}\propto M_\star^{2/5}$, given that $M_\star\propto v_{\rm h}^5$.
In \ikea\,, {\em the origin of the mass-metallicity relation is the dependence of the star formation efficiency on the halo's virial velocity.
The $M_\star-Z$  relation evolves because the $M_\star-v_{\rm h}$ relation evolves.} The first part of this claim agrees with \cite{Dave12}, but the second part does not: in their model, evolution is caused by the increase in metallicity of accreting gas. Note that, as long as the galaxy self-regulates\footnote{Clearly this would not true in case of a recent merger which might increase $\dot M_\star$ {\em and} dilute $Z$ by gas accretion.} its gas metallicity is set by the instantaneous star formation rate rather than a consequence of the build-up of metals that fail to escape from the potential well of its host halo. In other words, the reason that $Z$ depends on $\vh$ is because $\dot M_\star/\dot M_{\rm h}$ depends on $v_{\rm h}$, rather than that it is \lq easier for metals to escape from halos with low $v_{\rm h}$\rq, as is often claimed. Indeed we have assumed that $Z_{\rm w}=Z$ so that an outflow by itself does not affect $Z$ {\em at all}. Instead, low $v_{\rm h}$ halos have galaxies with low $Z$ because they are inefficient at forming stars.

\subsection{When stellar feedback fails}
\label{sect:AGN}
\begin{figure}
	\centering
	\includegraphics[width=\linewidth]{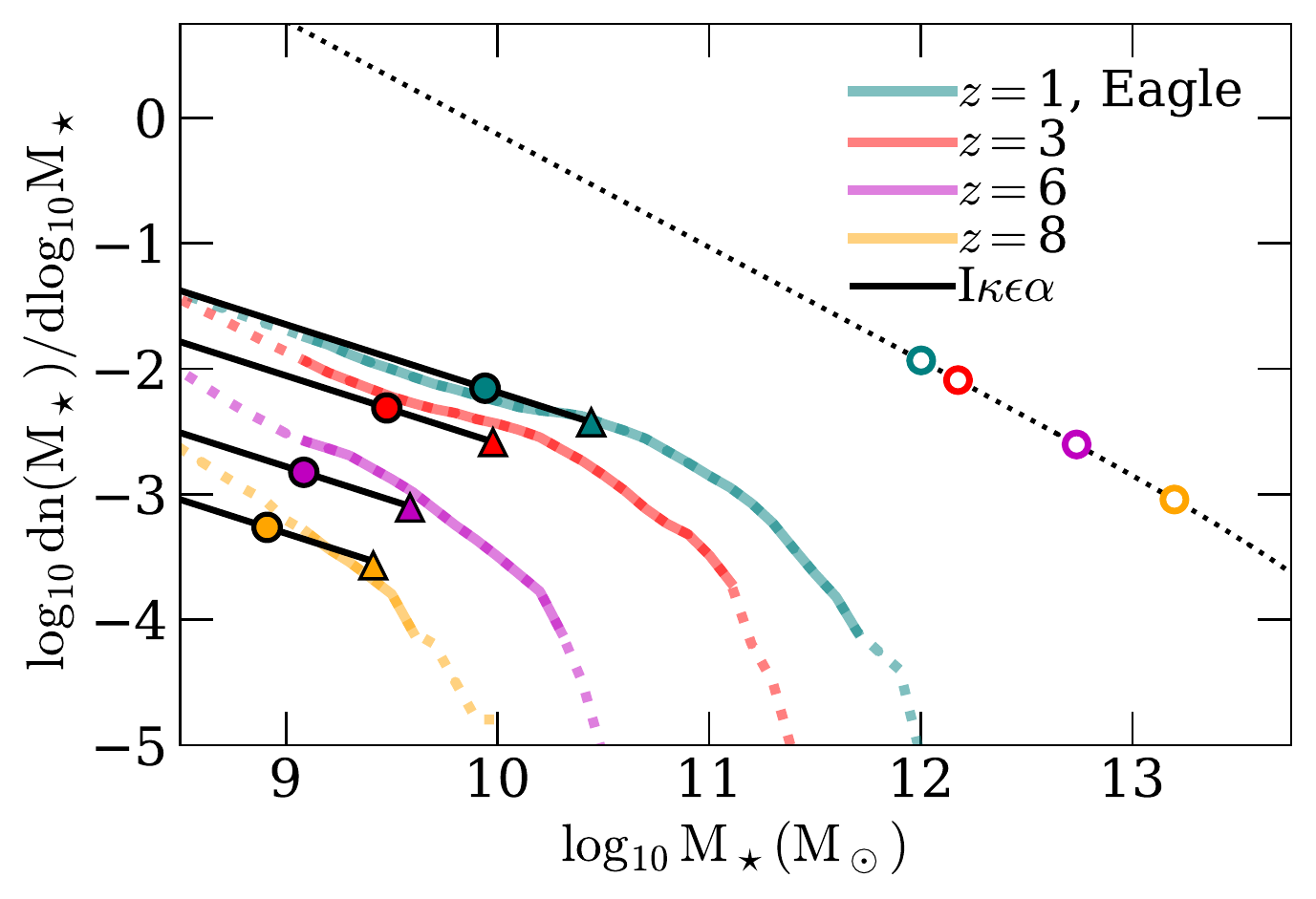}
	\caption{The evolution of the galaxy stellar mass function (GSMF). {\em Coloured curves} show the \eagle\ GSMF (simulation FbConst, in which the feedback efficiency is constant and which includes AGN feedback), with results at $z=1$, 3, 6, and 8 shown in {\em green}, {\em red}, {\em purple} and {\em yellow}, respectively. At high mass, curves are drawn as {\em dashed lines} if there are fewer than 5 galaxies dex in $\log_{10}M_\star$, at low mass when there are fewer than 100 stellar particles per galaxy. {\em Black full lines} are the corresponding \ikea\ results from \S \ref{sect:gsmf}, with a {\em triangle} corresponding to galaxies of mass $M_{\star, {\rm agn}}$  (Eq.~\ref{eq:Magn}) above which feedback from AGN is expected to set in, and a {\em filled circle} at half this mass. The {\em black dotted line} is the halo mass function, Eq.\ref{eq:hmf}. The {\em coloured open circles} indicate the abundance of halos with mass $M_{\rm h,{\rm agn}(z)}/2$, computed from Eq.~(\ref{eq:Magn}). 	The \ikea\ model predicts the shape and evolution of the normalization of the \eagle\ GSMF well. The predicted location of the knee in the GSMF is also reasonable. We used $a=\bar a$ and $b=\bar b$ for the accretion history of halos, Eq.~(\ref{eq:correaMp}).}
	\label{fig:GSMF}
\end{figure}

The basic \ikea\ relation of Eq.~(\ref{eq:SR}) between the halo accretion rate and the star formation rate results in $\dot M_\star\propto v_{\rm h,0}^{5}$,
where $v_{\rm h,0}$ is the virial velocity of the halo at redshift $z=0$, so that halos with a large virial velocity form stars at a greater rate. For low values of $v_{\rm h,0}$, only a very small fraction of the accreted baryons are converted into stars with the majority of the accreted gas expelled in an outflow, as discussed in \S \ref{sect:outflows}. The rate of gas accretion increases $\propto v^3_{\rm h,0}$ but the star formation rate increases $\propto v_{\rm h,0}^{5}$. Since obviously the star formation rate cannot be higher than when {\em all} accreted gas is converted to stars, $\dot M_\star\le \omega_{\rm b}\dot\Mh$, it eventually becomes impossible to satisfy Eq.~(\ref{eq:SR}) when \vh\ is larger than the critical value that results from inserting $\dot M_\star= \omega_{\rm b}\dot\Mh$ in Eq.~(\ref{eq:SR}),
\begin{equation}
v_{\rm h, max} = {v_\star\over \kappa^{1/2}}=310 {v_\star/400~{\rm km}~{\rm s}^{-1}\over (\kappa/(5/3))^{1/2}}~{\rm km}~{\rm s}^{-1}\,.
\label{eq:vcrit}
\end{equation}
Equation~\ref{eq:MB2} gives a slightly weaker limit when requiring that the mass-loading factor $\beta\ge 0$ so that any outflow decreases the baryon fraction of the halo rather than spuriously increasing it.

As a halo grows and $\vh$ increases, $\beta$ starts to drop rapidly to values below $1$, as seen in Fig.~\ref{fig:beta}. At $z\le 2$, $\beta\rightarrow 0$ for $\vh\rightarrow v_{\rm h,{\rm max}}$ but $\beta$ already plunges to values $\beta\le 1$
as $\vh$ approaches a somewhat smaller critical velocity. At higher redshifts, this critical halo virial velocity decreases, basically because it is no longer a good approximation to neglect $\dot M_{\rm gas}$. 

What is the consequence of this failure of self-regulation for halos with too high $\vh$? The \ikea\ GSMF discussed in \S \ref{sect:gsmf} is a power-law that tracks the power-law shape of the halo mass function. In contrast, the {\em observed} GSMF has an exponential cut-off at stellar masses above a characteristic stellar mass. It is thought that feedback from accreting black holes (AGN) suppresses star formation in such massive galaxies and this is the cause of the observed break in the GSMF \citep[e.g.][]{Bower06,Croton06}.

This motivates us to associate the critical velocity above which stars cannot self-regulate galaxy formation with those halos in which
AGN regulate galaxy formation instead. Using the subscript \lq agn\rq\ as a mnemonic, we see from Fig.~\ref{fig:beta} that the onset of AGN activity takes place at a nearly redshift-independent value of \vh\ of order
\begin{equation}
v_{\rm h, {\rm agn}}\approx 180~{\rm km}~{\rm s}^{-1}\,,
\end{equation}
for which the corresponding virial temperature is\footnote{$m_p$ is the proton mass, $k_{\rm B}$ is Boltzmann's constant and $\mu$ the mean molecular weight.}
\begin{equation}
T_{\rm h,\rm agn} = {\mu m_p v_{\rm h, {\rm agn}}^2\over 5{\rm k_{\rm B}}}
	\approx 10^{5.7}\,{\mu\over 0.62}\,
	\left({v_{\rm h, {\rm agn}}\over 180~{\rm km}~{\rm s}^{-1}}\right)^2~{\rm K}\,.
	\label{eq:Tvir}
\end{equation}

In the model described by \cite{Bower17}, seed black holes start to grow exponentially in mass when the outflow that is powered by feedback from star formation ceases to be buoyant in the hot corona that fills the dark matter halo. This causes a build-up of gas that fuels the growth of the black hole. The episode of exponential growth ends when the black hole is sufficiently massive that its feedback regulates the forming galaxy. In practise this results in a significant decrease in $\dot M_\star/M_\star$. This model describes well the behaviour of galaxies in the \eagle\ simulation, with the transition between star formation and AGN feedback regulated galaxies occurring in halos with a virial temperature nearly identical to that of Eq.~(\ref{eq:Tvir}) \citep{Mcalpline18}.

At first sight it seems that the reasoning that led to Eq.~(\ref{eq:Tvir}), \lq stellar feedback fails because $v_\star^2$, a measure of the thermal energy of feedback-heated gas, is too low compared to $\kappa\vh^2$\rq\, is very different from that of \cite{Bower17}, \lq stellar feedback fails because outflows are no longer buoyant in the hot corona\rq.  However, the build-up of the hot halo is itself depending on the efficiency of stellar feedback \citep{correa18}. Put in terms of \ikea: the higher $\epsilon$, the higher the value of \vh\ above which a hot corona develops (see in particular Fig.~14 in \citealt{correa18}). Within the current interpretation, the failure of stellar feedback is not due to the formation of a hot corona, but rather the formation of a hot halo is facilitated by failing stellar feedback. 

The results from previous sections allow us to compute other properties of the halo and the galaxy when $\vh=v_{\rm h, {\rm agn}}$, the onset of AGN activity. The halo mass, stellar mass, and star formation rate in a halo with $\vh=v_{\rm h, {\rm agn}}$ at $z=0$, are

\begin{eqnarray}
M_{\rm h,{\rm agn}}(z=0)&=& 2\times 10^{12}{\rm M}_\odot {(v_{\rm h,{\rm agn}}/180{\rm km}~{\rm s}^{-1})^3\over \alpha/1}\nonumber\\
M_{\star, {\rm agn}}(z=0) &=& 5.3\times10^{10} {\rm M_\odot} \left({v_{\rm h,{\rm agn}}\over 180{\rm km}~{\rm s}^{-1}}\right)^5\nonumber\\
\dot M_{\star, {\rm agn}}(z=0) &=& 	3.8\, {\rm M}_\odot{\rm yr}^{-1} \left({v_{\rm h,{\rm agn}}\over 180{\rm km}~{\rm s}^{-1}}\right)^5\,,
\end{eqnarray}
and the corresponding values at redshift $z$ are

\begin{eqnarray}
M_{\rm h,{\rm agn}}(z) &=& {M_{\rm h,{\rm agn}}(z=0)\over \Hub(z)}\nonumber\\
M_{\star,{\rm agn}}(z) &=& M_{\star,{\rm agn}}(z=0)\,{m_\star(z)\over (m_{\rm h}(z)\Hub(z))^{5/3}}\nonumber\\
\dot M_{\star, {\rm agn}}(z) &=& \dot M_{\star, {\rm agn}}(z=0) \,{\Psi_\star(z)\over (m_{\rm h}(z)\Hub(z))^{5/3}}\,.\nonumber\\
\label{eq:Magn}
\end{eqnarray}

We do not expect the \ikea\ GSMF to be correct for halos with $\vh\ge v_{\rm h, {\rm agn}}$. We therefore plot the GSMF discussed in
\S \ref{sect:gsmf} up to halos of mass $M_{\star,{\rm agn}}(z)$, and compare to the \eagle\ GSMF (simulation FbConst, in which the stellar feedback efficiency is a constant and which {\em does} include feedback from AGN) in Fig.~\ref{fig:GSMF}. The \ikea\ model reproduces the power-law shape of the \eagle\ mass function up to $M_{\rm h,{\rm agn}}(z)$ well, getting the evolution of the normalisation approximately correct as well. The value of $M_{\star,{\rm agn}}(z)$ is close to where \eagle\ predicts a rapid decrease in the number density of galaxies, which is due to the action of AGN feedback in the simulation. The number density of galaxies at the knee decreases with increasing $z$. The previous equations elucidate the reason for this in \ikea. Consider two redshifts $z_1$ and $z_2$, with $z_1<z_2$, say. Haloes with $\vh=v_{\rm h,{\rm agn}}$ at a redshift $z_2$ will be more massive at $z=0$ than those that have $\vh=v_{\rm h,{\rm agn}}$ at a redshift $z_1$, by the factor $\Hub(z_2)/\Hub(z_1)$, which is $\approx \left((1+z_2)/(1+z_1)\right)^{3/2}$ for $z_1\ge 1$. The corresponding ratio of number densities then follows from the slope of the PS halo-mass function, $\left(\Hub(z_2)/\Hub(z_1)\right)^{\alpha_{\rm h}}$. For example the co-moving number density at $z=6$ is lower than at $z=1$ by a factor 4.8.

\subsection{Reality check}
\label{sec:real}
Up to now we have compared \ikea\ to an \eagle\ simulation in which the feedback parameters are kept constant (simulation FbConst). That simulation does not reproduce the observed properties as well as the \eagle\ reference simulation. So, how well does \ikea\ describe the observations?

For a fiducial value of $v_\star=400~{\rm km}~{\rm s}^{-1}$, \ikea\ predicts that a $z=0$ galaxy with stellar mass $M_\star=5\times 10^{10}{\rm M}_\odot$ has a star formation rate of $\dot M_\star=3.5~{\rm M}_\odot~{\rm yr}^{-1}$ and is hosted in a dark matter halo of mass $M_{\rm h}=1.9\times 10^{12}{\rm M}_\odot$. For the Milky Way, the inferred values are $M_\star=(5\pm 1)\times 10^{10}{\rm M}_\odot$, $\dot M_\star=(1.65\pm 0.19){\rm M}_\odot~{\rm yr}^{-1}$ and $\Mh=(1.1\pm0.3)\times 10^{12}{\rm M}_\odot$ \citep{BlandHawthorn16}, respectively. However, the scatter in $M_\star$ and $\dot M_\star$ for a halo with given $M_{\rm h}$ is substantial, and the \ikea\ value for $M_\star$ is consistent with the abundance matching analysis by \cite{Guo10} and the star formation rate of $\dot M_\star=3.5~{\rm M}_\odot~{\rm yr}^{-1}$ falls well within the blue cloud for a galaxy with that $M_\star$ in the MPA-JHU DR7\footnote{\url{https://wwwmpa.mpa-garching.mpg.de/SDSS/DR7/}} catalogue. This reasonable level of agreement is of course not surprising: we chose \ikea's feedback efficiency parameter $\epsilon$ which sets $v_\star$ by comparing to these data sets.

The \ikea\ specific star formation rate is $\dot M_\star/M_\star\approx  0.07~{\rm Gyr}^{-1}$ at $z=0$, {\em independent of $\epsilon$}, as compared to an observed value of 0.1~Gyr$^{-1}$ at $M_\star=10^{10}{\rm M}_\odot$ (see the discussion of the data compilation by \citealt{Behroozi18}).  The observed sSFR increases to a value of $1~{\rm Gyr}^{-1}$ ($2~{\rm Gyr}^{-1}$) by redshift $z=1$ ($z=2$,  \citealt{Behroozi18}), as compared to the \ikea\ values of 0.3 (1). The \ikea\ values are actually very close to those in \eagle\ (simulation FbConstNoAGN). The faster observed evolution might signal that $\epsilon$ does evolve.

The  $M_\star\propto M_{\rm h}^{5/3}$ dependence of stellar mass on halo mass according to Eq.~(\ref{eq:mstar}) results in a redshift-independent low-mass slope of the galaxy stellar mass function of $dn/d\log(M_\star)\propto M_\star^{\approx -0.54}$. The faint-end slope of the Schechter luminosity function \citep{Schechter76},
\begin{equation}
{dn(L)\over d\log L}\propto L^{-\alpha_{\rm g}}\,\exp(-L/L_\star)\,,
\label{eq:Shechter}
\end{equation}
is $\alpha_{\rm g}\approx 0.48$ at redshift $z=0$ in the {\sc gama} \lq z\rq-band \citep{Loveday12}, a long enough wavelength so that stellar mass is approximately proportional to z-band luminosity. The level of agreement between the two slopes, 0.54 versus 0.48, is encouraging, but not surprising given that \ikea\ reproduces the \eagle\ GSMF at the low mass-end well (Fig.~\ref{fig:GSMF}). Observationally there is no convincing evolution of this slope out to $z\sim 3$ in the $K$-band \citep{Mortlock17}, also consistent with the \ikea\ prediction of no evolution. 

The observed evolution in the location of the knee of the Schechter luminosity function is claimed to be consistent with little or no evolution in the value of the {\em stellar mass} at which the transition occurs  (e.g. \citealt{Song16}) but an alternative interpretation is that the transition occurs at a nearly constant star formation rate. Indeed, according to \cite{Parsa16}, the absolute 1500\AA\ magnitude of galaxies at the knee of the Schechter luminosity function occurs at $M_{1500,c}=-19.6$, -20.3, -20.6 and -20.68 for redshifts $z=1$, 2, 3, 4, respectively. If we make the reasonable assumption that UV-luminosity is proportional to star formation rate, then the star formation rate $\dot M_{\star}$ of those galaxies increases compared to the value at $z=1$ by factors  $\dot M_{\star}(z)/\dot M_{\star}(z=1)=1.9$, 2.5, 2.8 and 2.9 at $z=2$, 3, 4 and 5. The prediction from \ikea\ follows from Eq.~(\ref{eq:Magn}), $\dot M_{\star,{\rm agn}}(z)/\dot M_{\star,{\rm agn}}(z=1)=1.6$, 2.2, 2.8 and 3.4, respectively, impressively close to the observations.

We conclude from this brief comparison to data that \ikea\ reproduces observations of the observed galaxy population and its evolution rather well, although there are some differences too.

\subsection{Incorporating AGN feedback}
\begin{figure}
	\centering
	\includegraphics[width=\linewidth]{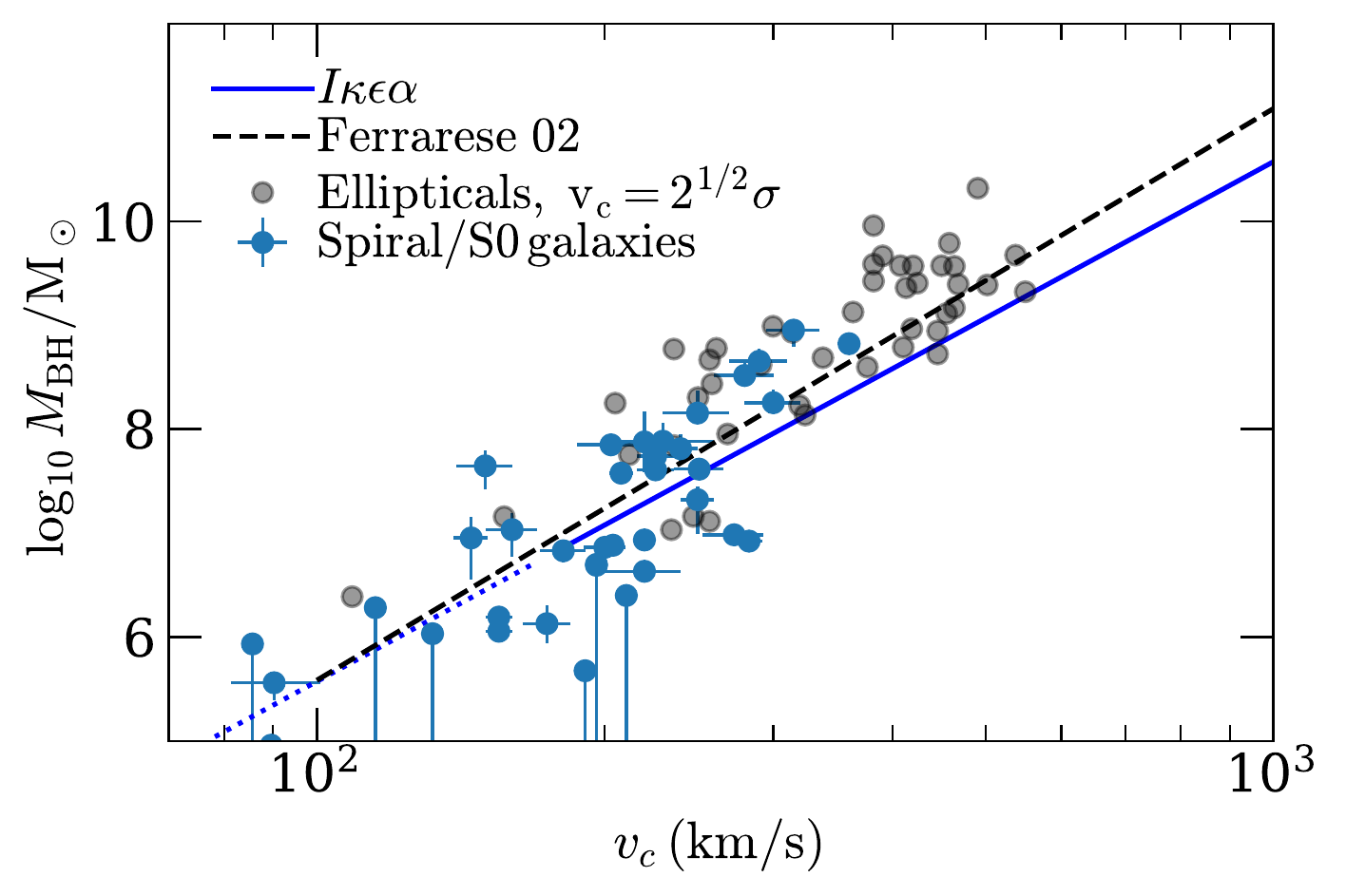}
	\caption{The dependence of black hole mass, $M_{\rm BH}$, on the circular speed of the host halo, $v_c$. The {\it blue line} is the trend from \ikea, Eq.~(\ref{eq:BH}), with $v_{\rm h}$ scaled  to $v_c$ following \citealt{Ferrarese02}; the trend originally proposed by \citet{Ferrarese02} is shown as a {\it dashed black line}. Data points are taken from the compilation by \citet{Kormendy13}; {\it blue circles} with error bars are spiral and S0 galaxies and {\it grey circles} are elliptical galaxies. }
	\label{fig:BH}
\end{figure}
An important limitation of the model as described so far is the absence of AGN feedback. Following the arguments that led us to stellar-feedback self-regulation,
an obvious way to include AGN in the model is by modifying Eq.~(\ref{eq:SR}) to
\begin{eqnarray}
{1\over 2}\dot M_\star\,v_\star^2 + {1\over 2}\dot M_{\rm BH}\,v_{\rm agn}^2&=& {\kappa\over 2}\omega_{\rm b}\,\dMh\vh^2\nonumber\\
v_{\rm agn}^2 &=&{2\epsilon_r\epsilon_f\over 1-\epsilon_r} c^2\,,
\label{eq:vbh}
\end{eqnarray}
with the understanding that AGN feedback sets in when\footnote{$v_{\rm agn}$, which characterises the energy input by the AGN per unit of mass accreted onto the BH, is not to be confused with $v_{\rm h, {\rm agn}}$ - the virial velocity of the halo above which stellar feedback fails.} $\vh\gtrapprox v_{\rm h,{\rm agn}}$. Here, $\epsilon_r\approx 0.1$ is the radiative efficiency of the AGN and $\epsilon_f\approx 0.15$ the fraction of radiated energy that couples to the gas (see the discussion in  \citealt{Schaye15}, their section 4.6).  As in the case of feedback from star formation discussed in \S \ref{sect:selfreg}, AGN feedback will be self-regulating provided that the black hole accretion rate {\em increases} with the pressure in the ISM.

We can examine what to expect for the black hole mass of a halo with $\vh\gg v_{\rm h,{\rm agn}}$ by
integrating Eq.~(\ref{eq:vbh}) for $v_\star=0$, 
\begin{eqnarray}
M_{\rm BH} &=& {3\over 5}{\kappa\omega_{\rm b}\over \alpha^{3/2}}  {v_{\rm h}^5\over 10H(z){\rm G}v_{\rm agn}^2}\nonumber\\
 &=& 1.2\times 10^7\,{(v_{\rm h}/200~{\rm km}~{\rm s}^{-1})^5\over (\epsilon_r\epsilon_f/(1-\epsilon_r))/(0.1\times 0.07)}{\rm M}_\odot\,,\nonumber\\
\label{eq:BH}
\end{eqnarray}
where the second line is at redshift $z=0$; this scaling is plotted as a blue line in Fig.\ref{fig:BH}. How does this compare to observations? \citet{Ferrarese02} claim that black hole mass scales with the circular speed as $M_{\rm BH} \propto v_{c}^{5.5}$. This scaling is shown as a black dashed line and is close to Eq.~(\ref{eq:BH}).  \citet{Kormendy13} argue that, because the scatter in the $M_{\rm BH}\hbox{--}v_c$ relation is large at low $v_c$, the  {\it Magorrian relation} \citep{Magorrian98} between black hole mass and bulge mass, is more fundamental. We would argue instead that low-mass black holes are not in a self-regulating regime. 

The observed relation between black hole mass and (3D)\footnote{We have assumed that $\sigma_\star^2=3\sigma^2$, where $\sigma$ is the line-of-sight stellar velocity dispersion.} stellar velocity dispersion $\sigma_\star$, is  $M_{\rm BH}=1.1\times 10^7\,(\sigma_\star/200~{\rm km}~{\rm s}^{-1})^{5.12}~{\rm M}_\odot$ \citep{McConnell11}. Provided $\vh\sim \sigma_\star$, the observed dependence on velocity is close to our prediction while the normalisation requires reasonable values for $\epsilon_r$ and $\epsilon_f$.  The scaling of the $M_{\rm BH}-\sigma_\star$ relation in the model by \cite{Silk98} is identical to ours basically because both are based on energy arguments, however, our normalisation is significantly more realistic, as shown in Fig.~\ref{fig:BH}. The model by \cite{King03} is based on momentum arguments; their scaling, $M_{\rm BH}\propto \sigma_\star^4$,  is shallower than observed. \cite{Booth10} obtain a
$M_{\rm BH}\propto M_{\rm h}^{1.55}$ scaling by arguing that the net total energy injected by an AGN is of order of the binding energy of a halo.
This is somewhat similar to our reasoning, except that we argue that it is the {\em rate} of energy injection by the AGN that tracks the {\em rate} of energy accretion by the halo due to self-regulation. The secular growth rate of a black hole - and hence the time-averaged luminosity of the AGN - therefore depends on the cosmological accretion rate onto its host halo and therefore on redshift, and not just on halo properties.

\section{Discussion}
\subsection{Comparison to previous work}
The paper by \cite{Bouche10} sparked interest in trying to understand the basic physics underlying self-regulation of galaxies. That paper, and several that followed, contain equations that resemble those of Section~3 - but the underlying assumptions are sometimes strikingly different, as we discuss here. The starting point of \cite{Bouche10} is their realisation that the dependence of $\dot M_\star$ on 
	stellar mass and redshift, resembles that of the cosmological accretion rate, suggesting that the
	gas accretion rate $\dot M_{\rm gas, acc}\propto \dot M_{\rm h}$. The proportionality constant is argued
	to be less than $\omega_b$, the cosmological gas to total matter density, because only cold accreted gas is assumed to be eligible for star formation. The resulting star formation rate, is then determined by the efficiency with which gas is converted into stars - that is - by the star formation law.

This reasoning results in $\dot M_\star\propto \omega_b\dot M_{h}$, as in our Eq.~(\ref{eq:dotmstar}), with the important distinction that the efficiency of galaxy formation,
\begin{equation}
  \label{eq:epsgal}
  \epsilon_g\equiv {\dot M_\star\over \omega_b\dot M_{h}}\,,
\end{equation}
is set by the efficiency of star formation, 
\begin{equation}
\epsilon_\star\equiv {\dot M_\star\over M_{\rm gas}/\tau_d}\,,
\label{eq:epsg}
\end{equation}
where $\tau_d$ is a characteristic time that still needs to be determined. The onus of getting the observed $M_\star/\Mh$ relation is now wholly on the star formation law, Eq.~(\ref{eq:epsg}). The solution advocated by \cite{Bouche10} , is to assume that halos do not form any stars as long as their halo mass is below some minimum value, $M_{h, min}\approx 10^{10}-10^{11}{\rm M}_\odot$, which conspires to result in $\epsilon_g$ increasing with \Mh. They stress repeatedly that their results are completely independent of the efficiency of feedback.

\cite{Lilly13} build on this work, and in their \lq gas regulator\rq\ frame work,
	$\dot M_\star$ is regulated by the gas reservoir of the galaxy, $M_{\rm gas}$ in our notation. Rather than assuming a minimum halo mass $M_{h, min}$ below which no stars form, the model introduces two main fitting parameters, which in our notation are the product $\epsilon_\star\,\tau_d$ (their variable $\epsilon$) and $\beta$ (their variable $\lambda$). In the follow-up paper by \cite{Birrer14}, they show how the evolution of galaxies over cosmic time can be modelled well once $\epsilon$ and $\lambda$ are parameterised as functions of $M_\star$. Note that these cannot be independent of $M_\star$, since otherwise the ratio $M_\star/\Mh$ is constant as well, since a constant fraction of the accreted gas is converted into stars.
	
The \lq minimum bathtub\rq\ model described by \cite{Dekel14} has very similar ingredients, in that $\dot M_\star$ is also regulated by $M_{\rm gas}$ through the star formation law. These authors stress that many properties of galaxies follow from this model if it is assumed that the system is in a quasi-steady state, $\dot M_{\rm gas}=0$.
	
These models \lq self-regulate\rq\ in the sense that the star formation rate is determined by the gas mass by mass conservation, in our notation $\dot M_{\rm gas}=\omega_b\dot \Mh - (1-{\cal R}+\beta)\dot M_\star$ (Eq.~\ref{eq:MB}}), so that too much star formation depletes the gas reservoir which ultimately decreases $\dot M_\star$. Conversely too little star formation leads to a build-up of $M_{\rm gas}$, and through the star formation law, this increases $\dot M_\star$. A very nice feature of these models, in addition to prediction correctly the rapid increase in $\dot M_\star/M_\star$ with redshift because the gas accretion rate $\propto \dMh$, is that they correctly predict secondary parameter dependencies, for example the fact that galaxies that lie above the main sequence are more gas rich and more metal poor, see also \cite{Dayal13}.

What all these models have in common is that the star formation rate is set by the gas reservoir through the star formation law. The origin of that law is not discussed in detail, but presumably it results from a balance between cooling and heating from star formation, as originally envisioned by \cite{White91}.
	In these models, feedback from star formation is only important in setting the
	star formation law, basically parameterised by $\epsilon_\star$. Combined with a model for the build-up of dark matter halos, or using dark matter-only simulations that follow the growth of halos, these \lq self-regulation\rq\ models are very successful in building realistic looking mock universes \citep[see e.g.][]{Moster18, Tacchella18}.

In our opinion, there are two major weaknesses to this basic model: ({\em i}) to be predictive the model needs to be able to predict how the efficiency of star formation, $\epsilon_\star$, and the mass loading factor, $\beta$,  depend on halo (or stellar mass), a formidable task. More worryingly, ({\em ii}) there is evidence that one of the main assumptions - that the star formation rate depends on the gas mass through the star formation law - is not quite correct.

At first sight it seems impossible that the rate of star formation in a galaxy is not dependent on the star formation law - and in fact it would be if the galaxy were isolated. However a galaxy in a cosmological setting can gain mass through accretion and lose it through winds - and therefore the amount of gas in the reservoir is not some constant, rather $M_{\rm gas}$ too is set by the physics of galaxy formation. Demanding that $\dot M_\star$ depends on $M_{\rm gas}$  through a star formation law, and vice versa, results in a \lq chicken and egg\rq\ problem.

Numerical simulations can be very helpful in distinguishing cause from effect. The {\sc owls} simulations described by \cite{Schaye10} are cosmological hydrodynamical simulations performed with {\sc gadget} \citep{Springel05}, but the parameters of sub-grid models are varied over a very wide range and not calibrated to observations as in \eagle. In particular, the {\sc owls} simulation suite includes parameter variations in which the efficiency of feedback from stars ({\em i.e.} the value of $v_{\rm \star}$ in our notation) and the star formation law (the values of $A$ and $n$ in Eq.~\ref{eq:KS}), are varied separately. By plotting $M_{\rm gas}$ and $\dot M_\star$ versus a variable that does not depend on either $v_\star$ or the star formation law, such as halo mass, \Mh, it becomes possible to test the very core assumption of the gas-regulator or bath-tub models.

\cite{Haas13a} compares models with the same star formation law (same value of $A$ and $n$) but different values of the feedback efficiency. Compare in particular their models REF and WML4: these have identical numerical parameters, except that the value of $v_\star^2$ in simulation WML4 is twice that of REF. Maybe not surprisingly, $\dot M_\star/\Mh$ in the simulation with the stronger feedback is about half as large as in REF (their Fig.~4). Because the star formation law in these simulations is the same, this also implies that $M_{\rm gas}/\Mh$ is also approximately half in WLM4 compared to REF, as is also born out by the same figure.

However now compare models REF and SFAMPLx3 in \cite{Haas13b}: these have identical feedback parameters, but the value of $A$ (from Eq.~\ref{eq:KS}) in simulation SFAMPLx3 is three times that in simulation REF. Figure~5 in \cite{Haas13b} shows that nevertheless the ratio $\dot M_\star/\Mh$ is nearly identical in the two simulations: the star formation rate in a halo of given mass is not, or only very weakly dependent, on $A$: a direct violation of the main assumption in the \lq gas-regulator\rq\ models. Given that the star formation rates are the same in these models, but the star formation law differs, this must imply that the gas reservoir in SFAMPLx3 is less than that in REF at a given value of \Mh: the same figure~5 shows that indeed $M_{\rm gas}/\Mh$ is about a factor of three lower in model SFAMPLx3 compared to REF.
As stressed by \cite{Haas13b} and confirming what was found by \cite{Schaye10}: stellar feedback regulates the star formation rate by determining the amount of (star forming) gas. In this interpretation, {\em $\dot M_\star$ regulates $M_{\rm gas}$ through stellar feedback, rather than $M_{\rm gas}$ setting $\dot M_\star$ through a star formation law}.

The model presented by \cite{Dave12} incorporate self-regulation through feedback, as envisioned here. Because they limit their analysis to equilibrium states defined by $\dot M_{\rm gas}=0$, their results are actually very similar to the various incarnations of the bath-tub models. 

In our interpretation, self-regulation follows from energy conservation, Eq.~(\ref{eq:SR}), and in particular the fact that $\dot E_g=0$ is a secularly stable equilibrium (provided that  $\dot\rho_\star$ increases with pressure of the star forming gas). Therefore accretion sets the star formation rate, once the net energy input generated by forming stars is known. This sets the \lq efficiency of galaxy formation\rq\ (the ratio of the star formation rate over the cosmological baryon accretion rate onto a halo) to be
\begin{equation}
{\dot M_\star \over {\omega_b \dot M_{\rm h}}} = \kappa {v_{\rm h}^2\over v_\star^2}\,,
\end{equation}
which does not depend on the star formation law but on the properties of the halo (through \vh) and the efficiency of feedback (through $v_\star$)
This is in contrast to Eq.~(\ref{eq:epsg}).  The star formation law then determines the gas reservoir in the \ikea\ model, with any excess accreted gas expelled in a wind.

Combining the main \ikea\ relation between the star formation rate and the accretion rate on a halo of Eq.~(\ref{eq:SR}), with the relation between halo virial velocity and halo mass (Eq.~\ref{eq:vh}) and the equation for the growth of a halo (Eq.~\ref{eq:correaM}), allows us to write the star formation rate in terms of the halo accretion rate in the form

The relation between stellar mass and halo mass (Eq.~\ref{eq:mstar}) can be cast in the form
\begin{eqnarray}
\log_{10}\left({M_\star\over 10^{12}{\rm M}_\odot}\right) &=&{5\over 3}\log_{10}  \left({M_h\over 10^{12}{\rm M}_\odot}\right) + \log_{10} {\cal N}(z)\nonumber\\
\log_{10} {\cal N}(z) &=&\log_{10}\left({1.7\times 10^{10}{\rm M}_\odot\over 10^{12}{\rm M}_\odot} {1-{\cal R}\over 0.55} {m_\star(z)\over m_h(z)^{5/3}}\right)\,,\nonumber\\
\end{eqnarray}
which has the form of Eq.~(1) in the paper by \cite{Salcido19}, with their $\epsilon(M_h,z)=5/3$. These author show that a halo mass - stellar mass of this form can 
be integrated to give analytical relations for the galaxy stellar mass function and the evolution of the cosmic star formation rate density.

\subsection{Limitations of the model}
A forming galaxy can fail to be able to attain its equilibrium star formation rate given by Eq.~(\ref{eq:SR}) for several reasons. Consider for example what happens if $\dMh$ suddenly decreases - for example because the galaxy becomes a satellite. Star formation will nevertheless continue in accordance with the star formation law, depleting the gas reservoir. In such galaxies, the star formation rate is set by the gas consumption time scale, rather than regulated by feedback. A less extreme version of the same phenomenon occurs when $\dMh$ for a particular halo is unusually small compared to the ensemble average. The \ikea\ model does not correctly describe this situation and in particular is not applicable to satellite galaxies.

We have neglected the finite lifetimes of massive stars. We think this is unlikely to be a major limitation at lower redshifts when the dynamical time of any galaxy is {\em much} larger than the lifetimes of massive stars. However, the limitation may affect the onset of star formation in small galaxies at high redshift. When \vh\ is very low, gas cannot cool and our self-regulation argument will not correctly predict $\dot M_\star$. When the halo grows in mass it may pass the threshold where gas can cool on a short time scale, and star formation may be unable to self-regulate because of the finite lifetimes of massive stars. This may lead to a star burst which \ikea\ does not model correctly.

Not unrelated is what happens at high values of \vh\ at low redshift. The \ikea\ model predicts that feedback becomes inefficient for
$\vh\approx 180$~km~s$^{-1}$ following similar reasoning to \cite{Bower17}. We argued, as did \cite{Bower17}, that the resulting increase in gas mass triggers the AGN, which, once the black hole mass has increased sufficiently, will regulate the galaxy. However, by construction this occurs in the same halos that develop a hot halo of gas, so that it becomes unlikely that the right hand side of Eq.~(\ref{eq:SR}) describes correctly the rate at which gas enters the galaxy: it may simply add to the hot halo instead (see the discussion in \citet{Bouche10} on hot versus cold accretion).  We think therefore that it is unlikely that \ikea\ models such galaxies accurately. Moreover, galaxy-galaxy mergers contribute significantly to the mass growth of such galaxies, and we have not attempted to include these in the model either. 

We also neglected that stars may form from gas lost by previous generations of stars - such recycling may affect the star formation rate of galaxies at late times
when their stellar masses are high but the cosmological accretion rate low  \citep[e.g.][]{Oppenheimer10, Voort17}.  Gas lost from galaxies by winds may re-accrete later - again we have neglected this effect. More in general, we have neglected the possibility that the accretion rate differs from $\omega_b\dMh$. 

If Eq.~(\ref{eq:SR}) is indeed applicable, then it might be possible to estimate the scatter around the main sequence of star forming galaxies from the scatter of \dMh\ around the ensemble average. This would provide a good test of the basic assumption in our model.

\section{Summary and conclusions}
We have presented a model for star formation in galaxies that is motivated by the origin of the stability of nuclear fusion in main sequence (MS) stars. The energy generated by nuclear fusion in a MS star equals the rate at which energy is lost through radiation. This equilibrium is secularly stable because if the star loses energy, it heats up, which increases the rate at which fusion occurs. The analogy with a star forming galaxy is that the rate of energy injection by supernovae (and winds from their massive progenitor stars) equals the rate at which energy is lost due to cosmological accretion. This equilibrium is stable provided the star formation rate increases with the pressure of the star forming gas.

Equation~(\ref{eq:SR}), $(1/2)\dot M_\star v_\star^2=(\kappa/2)\omega_b\dMh\,\vh^2$, encapsulates this energy balance. Here, $v_\star^2$ is a measure of the effective energy injected per unit mass of star formed by feedback, so that the left hand side is the rate at which feedback increases the galaxy's energy. The right hand side of the equation is the energy loss term due to cosmological accretion ($\omega_b$ is the cosmological baryon to total mass fraction), with $\vh^2$ a measure of the depth of the dark halo's potential. In our \lq \ikea\rq\ model, the star formation rate is set by the cosmological accretion rate by energy balance. The predicted dependence of $\dot M_\star$ on redshift and virial velocity, \vh, or halo mass, \Mh, agrees very well with that measured in the \eagle\ cosmological hydrodynamical simulation \citep{Schaye15}, as shown in Figs. \ref{fig:exp1} and \ref{fig:exp2},  respectively.

The \ikea\ model has four parameters (I, $\kappa$, $\epsilon $, and $\alpha$; hence the name), which together shape the star forming sequence of galaxies. The parameter \lq I\rq\ stands for the (stellar) {\bf I}nitial mass function (IMF), which sets how much energy is available for feedback from star formation, in particular from the supernovae (SN) associated with star formation, as well as the recycled fraction ${\cal R}$ that relates the time integral of star formation to the stellar mass formed.  We have kept the IMF constant in this paper. The dimensionless parameters $\kappa$ and $\alpha$ quantify the rate of cosmological accretion onto a halo ($\kappa$), and the concentration of such halos ($\alpha$, see Eq.~\ref{eq:halo}). We find that $\kappa\approx 5/3$ and $\alpha\approx 1$, and have kept these constant as well.

We think that the main numerical parameter that affects our results is $\epsilon$, which is a measure of the fraction of the energy that is injected by SNe that effectively increases the energy of the star forming gas, rather than being radiated away. It relates $v_\star^2$ to the energy produced by SNe per unit mass (or more generally to the energy injected in the ISM as a result of recent star formation), see Eq.~(\ref{eq:sn}). If feedback is efficient, $\epsilon$ is large, and $\dot M_\star$ is small. The \eagle\ simulation has a parameter, $f_{\rm th}$, that controls what fraction of the available supernova energy is injected into the star forming gas. This means that $f_{\rm th}\approx \epsilon$, provided radiative loses are small. Because feedback {\em is} efficient\footnote{Gas heated by SNe has its temperature increased by $\Delta T=10^7$~K where its cooling rate is minimal and mostly independent of metallicity.}  in \eagle, radiative loses in SN-heated gas are mostly small, which explains why the \ikea\ model reproduces the \eagle\ model with $f_{\rm th}$ held constant relatively well. However, in the \eagle\ \reference\ model, $f_{\rm th}$ is allowed to vary as a function of density and metallicity in a way that is calibrated so that the simulation reproduces (some) observations. Therefore to improve the agreement of \ikea\ with data, we would need to understand how radiative loses depend on the interstellar medium of a star forming galaxy. It seems unlikely that there is a simple way to do so.

A striking feature of the model is that $\dot M_\star$ does not depend on the gas mass, $M_{\rm gas}$, unlike what is assumed in many self-regulating models \citep[e.g.][]{Bouche10, Lilly13}. We use a star formation law (in our case the Kennicutt-Schmidt law, \citealt{Kennicutt98}) to infer $M_{\rm gas}$ from $\dot M_\star$ - rather than the other way around. Doing so allows us to reproduce the $M_{\rm gas}-\dot M_\star$ relation in \eagle\, (Fig.~\ref{fig:Mgas}) as well as the mass-metallicity relation (Fig.~\ref{fig:Zvh}).

We tried to incorporate feedback from accreting black holes (AGN) by ({\em i}) identifying when feedback from star formation fails so that a black hole can grow, and ({\em ii}) include AGN in the self-regulation process. Stellar feedback fails in galaxies with deep enough potential wells, so that 
energy injected by stars cannot compensate for energy lost through accretion even if {\em all} accreted gas is converted into stars. We showed that this occurs in halos with virial velocity above a nearly redshift independent critical value of $\sim 180~{\rm km}~{\rm s}^{-1}$. Demanding that the AGN regulates galaxy formation results in a relation between the black hole mass and the virial velocity of the halo of the form $M_{\rm BH}\propto \vh^5$, which closely follows the observed relation.

In the Introduction we discussed how gas cooling is thought to play an important role in determining the rate at which a galaxy forms stars, to the extent that it may even be the main property that determines the location of the peak in the redshift evolution of the star formation rate density of the Universe  \citep{Hernquist03}. Numerical simulations at first sight support this claim directly: a simulation where the contribution from metals is not included when calculating the cooling rate - and hence where the cooling rate is lower - yields lower values of $M_\star/\Mh$ than when metals are included (compare models {\sc NOZCOOL} and model {\sc REF} in Fig.3 of \citealt{Haas13a}). However, a lower metallicity of star forming gas reduces cooling loses of injected feedback energy, increasing $\epsilon$ and hence reducing $\dot M_\star$: that sequence of events is also consistent with the findings from \citealt{Haas13a}). The main impact of metallicity on the cooling rate of the gas may be on the efficiency of feedback, rather than on the accretion rate. Of course this argument breaks down in halos where the virial temperature is so high that most of the gas is and remains hot.

We think that \ikea\ provides a simple way of calculating the properties of a galaxy in terms of those of its host halo - and the results so obtained agree reasonably well with those from much more sophisticated models and importantly also with data. We suggest that a better description of how cooling losses depend on the properties of a galaxy through its history would improve the quality of the theoretical prediction.

\section*{Acknowledgements}
We thank our colleagues (J. Schaye, M. Schaller, R. Crain and R. Bower) for sharing with us the data from the \eagle\ simulation; and we are grateful to L. Heck and J. Helly for providing the computing support. This study was funded by the Science and Technology Facilities Council [grant number ST/F001166/1]. The study made use of the DiRAC Data Centric system at Durham University which is run by the Institute for Computational Cosmology on behalf of the STFC DiRAC HPC Facility (www.dirac.ac.uk); the equipment was funded by BIS National E-Infrastructure capital grant ST/K00042X/1, STFC capital grant ST/H008519/1, and STFC DiRAC; as a part of the National E-Infrastructure. This research was also supported by the Australian Research Council Centre of Excellence for All Sky Astrophysics in 3 Dimensions (ASTRO 3D), through project number CE170100013. The International Centre for Radio Astronomy Research (ICRAR) is a Joint Venture of Curtin University and The University of Western Australia, funded by the Western Australian State government. M.S. is supported by an ASTRO 3D fellowship at the ICRAR, Curtin University.

\bibliographystyle{mnras}
\bibliography{self-regulation}
\appendix
\section{Simulation details}
We compare the results of the model described in this paper to galaxies from the \eagle\ simulation, which we briefly describe here. \lq Evolution and Assembly of Galaxies and their Environments\rq\  (\eagle) is a suite of cosmological, hydrodynamical simulations, performed using an evolution of the {\sc gadget} smoothed particle hydrodynamics code described by \cite{Springel05}. \eagle\ uses a set of sub-grid modules to encode unresolved physics, described in detail by \cite{Schaye15}, which we briefly summarise here.

The sub-grid modules contain a set of numerical parameters, whose values are calibrated in a {\sc reference} run to reproduce a small number of $z\approx 0$ observables, namely the galaxy stellar mass function, the relation between galaxy stellar mass, $M_\star$, and size, and between $M_\star$ and black hole mass, as detailed by \cite{Crain15}. Given these calibrated values, the simulation also reproduces several observable relations that were not part of the calibration, in particular yielding a \lq main sequence\rq\ of blue star forming galaxies in which $\dot M_\star$ depends on $M_\star$ and redshift as observed \citep{Furlong15}, as well as a \lq red sequence\rq\ of quenched galaxies \citep{Trayford15,Trayford17}. The $z=0$ galaxy colours correlate with galaxy morphology as observed \citep{Correa17,Trayford18}.
 
Most relevant for the comparisons in this paper are the implementation of star formation, of stellar feedback, and of feedback from accretion blackholes (AGN) in \eagle:
\begin{itemize}

\item {\em Star formation:} sufficiently dense gas in \eagle\ is converted into star particles at a rate per unit volume, $\dot\rho_\star$, that depends on the gas pressure, $P$, as 
\begin{equation}
	\label{eq:dotmstar1}
  \dot\rho_\star\propto P^{(n-1)/2}\,.
\end{equation}
The normalisation of this relation and the exponent $n$ are set by the Kennicutt-Schmidt law \citep{Kennicutt98} that relates the surface density of star formation, $\dot\Sigma_\star$ and of gas, $\Sigma_{\rm g}$, 
\begin{equation}
  \label{eq:KS}
  \dot\Sigma_\star = A \left({\Sigma_g\over 1~{\rm M}_\odot\,{\rm pc}^{-2}}\right)^{n}\,.
\end{equation}
The underlying assumption connecting these relations is that volume and surface densities are related by the local Jeans length, as motivated by \cite{Schaye08}. The simulation does not resolve the multi-phase nature of the interstellar medium and
	star forming gas is assumed to have a minimum pressure \citep{Schaye15},  
\begin{equation}
  \label{eq:eos}
  P \propto \rho^{4/3} \propto u^4\,,
\end{equation}
where $u$ is the thermal energy per unit mass.

\item {\em Stellar feedback} is implemented as described by \citet{DallaVecchia12}: a newly formed star particle increases the temperature of surrounding gas by an amount $\Delta T$. The quantity of gas heated depends on the effective energy injected by star formation, $f_{\rm th}\Delta E$, where $\Delta E$ is the total energy released by the winds from massive stars and core-collapse supernovae which in turn depends on the assumed stellar initial mass function (IMF). The value of $\Delta T$ is chosen such that gas is heated to a temperature where its cooling rate is small: this makes the feedback efficient. The value of $1-f_{\rm th}$ 	quantifies the fraction of injected energy that is lost from the star forming region, for example through radiative cooling; $f_{\rm th}$ is one of the main calibration parameters in \eagle.

\item {\em Black holes and AGN:} The seeding, merging, accretion, and feedback from black holes (BHs) in \eagle\ is described by \cite{Rosas15}. Seed BHs are inserted in each dark matter halo once it becomes sufficiently well resolved. When a BH accretes mass and becomes an AGN, it injects thermal energy in the surrounding gas. 
\end{itemize}

\begin{table} 
\begin{center}
\caption{Selected parameters of the \eagle\ simulations used here. From left-to-right the columns show: simulation name; co-moving box size; initial baryonic particle mass; maximum proper softening length, and comment.} 
\label{tbl:sims}
\begin{tabular}{lrrrrrrr}
\hline
Name & $L$ & $m_{\rm g}$ & $\epsilon_{\rm prop}$ & comment\\  
& (Mpc) & ($10^6 {\rm M}_\odot$) & (kpc)\\
\hline
REF          &  50 & $1.81$ & 0.7 & reference model \\  
FBconst      &  50 & $1.81$ & 0.7 & {\hbox{$f_{\rm th}=1$}}\\ 
FBconstnoAGN &  50 & $1.81$ & 0.7 & {\hbox{$f_{\rm th}=1$, no AGN}}\\
DMO          &  50 & 0 & 0.7 & dark matter only\\
\hline
\end{tabular}
\end{center}
\end{table}

The origin of red galaxies in \eagle\ is investigated by \cite{Trayford16}. Ram-pressure stripping and \lq strangulation\rq\ dramatically decreases the star formation rate of satellite galaxies, causing them to leave the blue cloud of star forming galaxies and settle onto the red sequence. The simple self-regulating model described in this paper does not attempt to describe these effects, and we will therefore only compare to central, {\em i.e.} non-satellite, \eagle\ galaxies. Similarly, AGN feedback suppresses star formation in massive galaxies, causing them to become passive. Since that mechanism is also not included in the model, most of the comparison in this paper i
to \eagle\ variation FbConstNoAGN, in which $f_{\rm th}$ is a constant, and which does not include AGN feedback.
We also use variation FbConst, in which $f_{\rm th}$ is kept constant and which {\em does} include AGN.

Table~\ref{tbl:sims} contains a list of parameters of the \eagle\ runs that we used. Simulation \lq REF\rq\ is the default \eagle\ model from Table~2 of \cite{Schaye15}. The simulation FBconst with $f_{\rm th}=1$ appears in Table~1 of \cite{Crain15},
	simulation. Simulation DMO is a dark matter-only version of the same volume. All simulations are initialised from the same Gaussian initial conditions, so that halo masses are nearly identical in all runs.

Galaxies of the \eagle\ reference model look like observed galaxies in many of their properties. Keeping $f_{\rm th}$ constant, the simulated galaxies have similar stellar masses and star formation rates, but are typically smaller than in the reference model. Therefore this model is not as good a representation of the real galaxy population, but we believe its physics is still reasonable - and it is much easier to compare to our simple model. Many of the properties of the population of \eagle\ galaxies can be extracted directly from the public database\footnote{\url{http://icc.dur.ac.uk/Eagle/database.php}.} \citep{McAlpine16}, which we used extensively in preparing the figures.

\end{document}